\documentclass[prb,preprint,eqsecnum,aps]{revtex4}
\usepackage{graphicx}

\newcommand{\ca}{{\cal A}}
\newcommand{\dt}{{\epsilon}}
\newcommand{\e}{{\rm e}}

\newcommand{\lm}{\lambda}

\newcommand{\ep}{\epsilon}

\newcommand{\ga}{\gamma}
\newcommand{\ka}{\kappa}

\newcommand{\za}{\zeta}
\newcommand{\sgn}{{\rm sgn}}
\newcommand{\p}{{\bf p}}
\newcommand{\x}{{\bf x}}
\newcommand{\bu}{{\bf u}}
\newcommand{\xp}{{\bf x}^\prime}
\newcommand{\rr}{{\bf R}}
\newcommand{\pp}{{\bf P}}
\newcommand{\br}{{\bf r}}
\newcommand{\s}{{\bf s}}

\newcommand{\dte}{{\epsilon}}

\newcommand{\bea}{\begin{eqnarray}}
\newcommand{\eea}{\end{eqnarray}}
\newcommand{\be}{\begin{equation}}
\newcommand{\ee}{\end{equation}}
\newcommand{\ba}{\begin{eqnarray}}
\newcommand{\ea}{\end{eqnarray}}

\newcommand{\nn}{\nonumber}
\newcommand{\la}{\label} 
\newcommand{\w}{\omega} 
 
\newcommand{\hT}{\hat{T}}
\newcommand{\hG}{\hat{G}}
\newcommand{\hV}{\hat{V}}

\newcommand{\pa}{\partial}

\begin{document}
	
\title{Understanding the sign problem from an exact Path Integral Monte Carlo model of
	interacting harmonic fermions}

\author{Siu A. Chin}
\email{chin@physics.tamu.edu.}

\affiliation{Department of Physics and Astronomy, 
	Texas A\&M University, College Station, TX 77843, USA}

\begin{abstract}
	
	This work shows that the recently discovered operator contraction identity
	for solving the discreet Path Integral of the harmonic oscillator
	can be applied equally to fermions in any dimension. This then yields an exactly solvable
	model for studying the sign problem where the Path Integral Monte Carlo energy at 
	any time step for any number of fermions is known analytically, or can be computed numerically.
	It is found that the sign problem is primarily a property of the free fermion propagator,
	but repulsive/attractive pairwise interaction can shift the sign problem to larger/smaller imaginary time
	but does not make it more severe than the non-interacting case. More surprisingly,
	one can prove analytically that the first closed-shell state in $D$ dimension, with
	$n=D+1$ fermion, has no sign problem at large imaginary time.
	Direct numerical simulations confirm that this is also true for higher 
	closed-shell states in two and three-dimension.
	Fourth-order and newly found variable-bead algorithms are used to compute ground
	state energies of quantum dots with up to 110 electrons and compared
	to results obtained by modern neural networks.

\end{abstract}


\maketitle

\section {Introduction}

Recent advances in Path Integral Monte Carlo (PIMC) continue to be the use of 
parameterized fourth-order propagators to optimize PIMC's convergence at the 
fewest number of beads/time-slices for both bosons\cite{jan01,sak09,zil10,kam16,kap16,yan17,lin18,wan22}
and fermions\cite{chin15,dor15,dor18}. While this increase in efficiency is 
useful for both, it is of crucial importance for fermions because its sign 
problem worsens with increasing number of beads. However, because there is no 
simple model for understanding PIMC for fermions at each time step, one's knowledge 
of the sign problem remains scant. For example, it is well known since Takahashi and Imada's 
calculation\cite{tak84a} 40 years ago that PIMC has no sign problem in one dimension. 
Yet, an explicit proof\cite{chin24} of this has only appeared recently. The
technique used in that proof can now be generalized to explain why there is no sign problem
for the first closed-shell particle states at large imaginary time. (See Sect.\ref{nosgn} below.)

In a previous work\cite{chin23a}, because of an operator contraction identity, this author
discovered the {\it universal} PIMC discrete propagator for the harmonic oscillator, which has
the same functional form for all initial short-time propagators. Each short-time propagator only modifies
the argument of the universal coefficient functions. This then makes it possible
to obtain physical observables in closed forms for all short-time propagators at once.
In particular, one can compute the thermodynamic and Hamiltonian energies analytically at 
each discrete imaginary time step for any short-time propagator\cite{chin23b}.

This work now shows that this contraction identity remains valid 
for the anti-symmetric determinant fermion propagator at all dimensions and for any number of fermions.
Therefore, one has a completely solvable PIMC model of harmonic fermions with analytical 
thermodynamic and Hamiltonian energies. 
Moreover, the model remains solvable if the pairwise interaction is also harmonic.
Therefore, the sign problem can be examined in detail, with and without interactions.

In Section \ref{oc}, we recall the contraction identity for the harmonic oscillator in one dimension,
then generalize it to any number of fermions in any dimension. By use of this contraction, 
Sect.\ref{upp} summaries the universal discrete PIMC propagator for harmonic fermions.
Sect.\ref{dpfr} derives the analytical form of the discrete partition function at each time-step
for all short-time propagators. 
Sect.\ref{athe} gives the analytical thermodynamic and Hamiltonian fermion energies 
at each discrete time-step verified by direct PIMC simulation.
Sect.\ref{sgpb} discusses the PIMC fermion sign problem and the empirical finding that there is no
sign problem for closed-shell particle states at large imaginary time.
Sect.\ref{nosgn} explains the surprising connection between the absence of the sign problem in 
one dimension and in closed-shell particle states.
The absence of the sign problem in $D$-dimension for the first closed-state with $n=D+1$ fermions
is proved in Appendix E.
Sect.\ref{sgint} shows how pairwise interactions, both harmonic and Coulombic, modify the sign problem from
the non-interacting case. 
For spin-balanced quantum dots, Sect.\ref{sbqd} shows that fourth-order algorithms give excellent
ground state energies for less than 30 electrons, provided that the Coulomb interaction is
sufficiently repulsive. Larger quantum dots necessitate the use of a new class of 
Variable-Bead algorithms in Sect.\ref{lqd}. The ensuing energies for quantum dots with up to 110 electrons 
are compared to modern neural network\cite{nor23} results. Conclusions and future directions are presented in Sect.\ref{con}.

\section {Free fermion propagator contractions}
\la{oc}

For the dimensionless harmonic oscillator Hamiltonian in 1D,
with kinetic and potential energy operators defined below 
\be
\hat H=-\frac12\frac{d^2}{dx^2}+\frac12 x^2\equiv \hat T+\hat V,
\ee
one has the operator identity\cite{chin23a}
\ba
\e^{ -a\hT}\e^{ -b\hV}\e^{-c\hT}
&=&\e^{ -\nu\hV}\e^{-\ka\hT}\e^{ -\mu\hV},
\la{oci}
\ea
where
\be
\ka=a+abc+c,\quad\nu=\frac{bc}{\ka},\quad\mu=\frac{ba}{\ka}.
\la{kauv}
\ee
Therefore any discrete path integral of the harmonic oscillator can be contracted
down to a single free propagator form and be evaluated analytically as
\ba
\langle x'|\e^{ -A\hV}\e^{-B\hT}\e^{ -C\hV}|x\rangle
=\e^{-A\frac12 {x'}^2}\frac1{\sqrt{2\pi B}}
\e^{-\frac1{2B}(x'-x)^2}\e^{-C\frac12 x^2}.
\la{gmat}
\ea

Consider the case of $n$-particles in $D$-dimension with configuration vector $\x=(\br_1,\br_2\cdots,\br_n)$ where
each particle position vector is $\br=(x_1,x_2,\cdots, x_D)$. Since the harmonic oscillator is separable in each coordinate,
the contraction (\ref{oci}) now holds for
\ba
\hat T=-\frac12\nabla_{\x}^2=-\frac12\sum_{i=1}^n\nabla_i^2
\qquad{\rm and}\qquad
\hat V=\frac12\x^2=\frac12\sum_{i=}^n\br_i^2
\ea
where the $n$-particle free propagator is given by
\ba
g_0(\x^\prime,\x,\dt)&=&\langle\x'|\e^{-\ep\hT}|\x\rangle\nn\\
&=&\prod_{i=1}^n \frac1{(2\pi\dt)^{D/2}}\exp\left[-\frac1{2\dt}(\br_i^\prime-\br_i)^2\right].
\ea

For $n$ fermions, the anti-symmetric free propagator $\e_\ca^{ -\dt\hT}$ in $D$ dimension is given by
the determinant
\ba
G_0(\x^\prime,\x,\dt)&=&
\langle \x^\prime|\e_\ca^{ -\dt\hT}|\x\rangle\nn\\
&=&\frac1{n!}{\rm det}\left(\frac1{(2\pi\dt)^{D/2}}
\exp\left[-\frac1{2\dt}(\br_i^\prime-\br_j)^2\right] \right)\la{gdet}\\
&=&	\frac1{n!}\sum_{p}(-1)^P g_0(\x^\prime_p,\x,\dt)\la{pone}\\
&=&	\frac1{n!}\sum_{p}(-1)^P g_0(\x^\prime,\x_p,\dt)\la{ptwo}
\ea
where 
$\x_p$ is a particle permutation of $\x$ with $P=0$ for even permutations and $P=1$ for
odd permutations. From from (\ref{gdet}), $G_0(\x^\prime,\x,\dt)$ is antisymmetric
in $\xp$ and in $\x$.

The contraction of two free fermion propagators gives
\ba
&&\langle \x^\prime|\e_\ca^{ -a\hT}\e^{ -b\hV}\e_\ca^{-c\hT}|\x^{*}\rangle
=\int d\x\, G_0(\x^\prime,\x,a)\e^{-b\frac12 \x^2}G_0(\x,\x^*,c)\nn\\
&&\qquad=\frac1{(n!)^2}\sum_{p,q}(-1)^{P+Q}\int d\x\, g_0(\x^\prime_p,\x,a)\e^{-b\frac12 \x^2}g_0(\x,\x^*_q,c)\nn\\
&&\qquad=\frac1{(n!)^2}\sum_{p,q}(-1)^{P+Q}\langle \x^\prime|\e^{ -a\hT}\e^{ -b\hV}\e^{-c\hT}|\x^{*}\rangle.
\ea
Since the contraction holds for the  $n$-particle free propagator, the above is just 
\ba
\langle \x^\prime|\e_\ca^{ -a\hT}\e^{ -b\hV}\e_\ca^{-c\hT}|\x^{*}\rangle
&&=\e^{-\nu\frac12 \x'^2-\mu\frac12 {\x^*}^2}
\frac1{(n!)^2}\sum_{p,q}(-1)^{P+Q}g_0(\x^\prime_p,\x^*_q,\ka)\nn\\
&&=\e^{-\nu\frac12 \x'^2-\mu\frac12 {\x^*}^2}
\frac1{n!}\sum_{p}(-1)^{P}G_0(\x^\prime_p,\x^*,\ka).
\ea
From (\ref{gdet}), any row permutation changes the sign according to 
\be
G_0(\x^\prime_p,\x,\dt)=(-1)^PG_0(\x^\prime,\x,\dt).
\la{schg}
\ee
Hence
\ba 
\langle \x^\prime|\e_\ca^{ -a\hT}\e^{ -b\hV}\e_\ca^{-c\hT}|\x^{*}\rangle
&=&\e^{-\nu\frac12 \x'^2-\mu\frac12 {\x^*}^2}G_0(\x^\prime,\x^*,\ka)\nn\\
&=&\langle \x^\prime|\e^{ -\nu\hV}\e_\ca^{-\ka\hT}\e^{ -\mu\hV}|\x^*\rangle.
\la{opid}
\ea
Thus we have shown that the operator contraction identity applies equally to free propagators 
of any number of harmonic fermions in any dimension.
This immediately implies that:
\begin{enumerate}
	
	\item[a)]
Since the sign of $G_0(\x^\prime,\x^*,\ka)$ depends only on permutations of particle positions 
and {\it not} on $\ka$, 
and contractions only change $\ka$, 
the nodal structure of the free fermion propagator is preserved under all contractions. 
Hence, the discrete path integral for non-interacting harmonic fermions at any imaginary time-step
has the {\it same} nodal structure as the free fermion propagator.

	\item[b)] For fermions confined by a general potential $V(\x)$, one can approximate
	\be
	\e^{-bV(\x)}=\sum_k c_k \e^{-b_k \x^2}
	\la{gtran}
	\ee
	by a sum of Gaussians. 
	The contraction then spawns multiple determinant propagators via 
	\be
	\int d\x\, G_0(\x^\prime,\x,a)\e^{-bV(\x)}G_0(\x,\x^*,c)
	=\sum_k c_k \e^{-\nu_k\frac12 \x'^2}G_0(\x^\prime,\x^*,\ka_k)\e^{-\mu_k\frac12 {\x^*}^2},
	\la{sumg}
	\ee
	and changes the nodal structure of the discrete path integral away from that of the free propagator. 
	For the pioneering approximation of Slater type orbital\cite{heh69}, no more than six Gaussians are needed.
	The approximation here will be practical if similar numbers of Gaussians are also suffice 
	in the expansion of (\ref{gtran}). 

	\item[c)] This contraction holds true for the determinant free fermion propagator (\ref{gdet})
	but is not true when one only samples permutations randomly in (\ref{pone}) or (\ref{ptwo}).
	Therefore, doing fermion PIMC via {\it permutation sampling}\cite{cep95,lyu05} will not precisely reproduce
	analytical results presented in this work.

	\end{enumerate}
Point a) explains why non-interacting harmonic fermions are as simple as free fermions. Point b) explains how generally
interacting fermions are complicated by multiple determinants. 
Point c) explains that the use of the determinant propagator, while computationally more intensive,
is superior to permutation sampling. This last point is related to an old finding by Lyubartsev\cite{lyu05},
that while there is no sign problem for 1D fermions when using the anti-symmetric propagator, there remains a low-level sign problem if one only does permutation sampling\cite{chin24}.
This work only uses the above anti-symmetric determinant free fermion propagator as originally
employed by Takahashi and Imada\cite{tak84a} and subsequent authors\cite{lyu05,chin15,dor15,chin24}.

\section{The universal discrete propagator of harmonic PIMC }
\la{upp}

Because of the contraction identity (\ref{opid}),
any left-right symmetric short-time operator of the form 
\be
\hG_1(\dt)=\prod_{i}\e_\ca^{ -a_i\dt\hT}\e^{ -b_i\dt\hV},
\ee 
can be contracted down to a single $\hT$-operator form
\be
\hG_1(\dt)=\e^{ -\mu_1\hV}\e_\ca^{ -\ka_1\hT}\e^{ -\mu_1\hV},
\la{sho}
\ee 
where $\mu_1$ and $\ka_1$ are functions of $a_i$, $b_i$, $\dt$
 and whose matrix element 
\ba 
G_1(\x^\prime,\x,\ep)=\langle \x^\prime|\hG_1(\dt)|\x\rangle
&=&\e^{-\mu_1\frac12 {\x'}^2}G_0(\x^\prime,\x,\ka_1)\e^{-\mu_1\frac12 {\x}^2}
\la{shp}
\ea
is the anti-symmetric, determinant short-time propagator. Since the  
the operator (\ref{sho}) is trivially related to its matrix element (\ref{shp}),
the propagator, we will use words ``operator" and ``propagator" intechangably. 
For the well-known second-order Primitive Approximation (PA) propagator, $\mu_1=\ep/2$ and $\ka_1=\ep$. 

Following the same steps as in Ref.\onlinecite{chin23a}, 
the contraction of $N$ short-time fermion propagator
$\hG_1(\dt)$ gives the discrete PIMC operator at $\tau=N\ep$
\be
\hG_N=[\hG_1(\dt)]^N=\e^{ -\mu_N\hV}\e_\ca^{ -\ka_N\hT}\e^{ -\mu_N\hV},
\la{gn}
\ee
with {\it universal} coefficients 
\ba
&&\za_N=\cosh(Nu),\la{hza} \\
&&\ka_N=\frac1{\ga}\sinh(Nu),\la{kka}\\
&&\mu_N=\frac{\za_N(Nu)-1}{\ka_N(Nu)}=\ga\tanh(Nu/2).
\la{mma}
\ea
Given $\ka_1$ and $\mu_1$ corresponding to $N=1$, the above equations
defines the $N$ time-step PIMC propagator (\ref{gn}) analytically.
In particular, (\ref{mma}) defines $\za_1=1+\ka_1\mu_1$ from which
the {\it portal} parameter $u$ can be deduced from (\ref{hza})
\be
u=\cosh^{-1}(\za_1)=\ln\left(\za_1(\ep)+\sqrt{\za_1^2(\ep)-1}\right).
\la{udef}
\ee
The coefficient $\ga$, defined at $N=1$ by (\ref{kka}), remains unchanged for all $N$:
\be
\ga=\frac{\sinh(u)}{\ka_1}=\frac{\sqrt{\za_1^2-1}}{\ka_1}=\frac{\sqrt{\za_N^2-1}}{\ka_N}.
\la{gai}
\ee
The discrete PIMC propagator (\ref{gn}) is a universal function of the portal parameter $u$
for {\it all} short-time propagators.
Each short-time propagator only changes $u$ through $\za_1$ via (\ref{udef}).

For PA, $\za_1=1+\dt^2/2$.
In the continuum limit of $\dt=\tau/N\rightarrow 0$ with $\tau$ fixed, 
\be
u=\cosh^{-1}(\za_1)\la{ucosh}\rightarrow\dt-\frac1{24}\dt^3+\cdots
\ee
\be
Nu \rightarrow N\dt=\tau,\qquad \ga=\sqrt{1+\ep^2/4}\rightarrow 1
\ee
and the PIMC propagator (\ref{gn}) becomes the exact $n$-harmonic-fermion propagator
with
\be
\ka_N\rightarrow\sinh(\tau),\la{kaa}
\ee
\be
\mu_N\rightarrow\frac{\cosh(\tau)-1}{\sinh(\tau)}.
\la{maa} 
\ee
That this exact propagator gives the correct ground state energy for up to 100 spin-polarized free fermions in a 
2D harmonic oscillator has been shown previously in Ref.\onlinecite{chin15}. The above discussion can be regarded
as its algorithmic derivation directly from PA. 

\section{Discrete fermion Partition functions}
\la{dpfr}

The partition function at $\tau=N\ep$ for $n$ fermion is given by
\ba
Z_n(\tau)&=&\int d\x G_N(\x,\x,\ep)\nn\\
&=&\frac1{n!}\frac1{(2\pi\ka_N)^{nD/2}}\int d\x\e^{-\mu_N{\x}^2}
{\rm det}\left( \exp\left[-\frac1{2\ka_N}(\br_i-\br_j)^2\right] \right)
\la{disz}
\ea

For $n=1$, one has
\ba
Z_1&=&\frac1{(2\pi\ka_N)^{D/2}}\int d\br_1\e^{-\mu_N {\br_1}^2}=\frac1{(2\pi\ka_N)^{D/2}}\left(\frac{\pi}{\mu_N}\right)^{D/2}\nn\\
&=& \left(\frac1{2(\za_N-1)}\right)^{D/2}=\left( \frac1{2 \sinh(w/2)}\right)^D
\ea
where we have defined
\be
w=Nu
\la{wdef}.
\ee 
This $w$ is the algorithm's imaginary time, in contrast to
the actual imaginary time of $\tau=N\ep$. They are equal in the continuum limit
of $u\rightarrow\ep$.

For $n=2$,
\ba
Z_2
&=&\frac1{2!}\frac1{(2\pi\ka_N)^{2D/2}}\int d\x\e^{-\mu_N{\x}^2}
{\rm det}\left( \exp\left[-\frac1{2\ka_N}(\br_i-\br_j)^2\right] \right)\nn\\
&=&\frac1{2!}\frac1{(2\pi\ka_N)^{2D/2}}\int d\br_1d\br_2\e^{-\mu_N(\br_1^2+\br_2^2)}
\left|\begin{array}{cc}
	1&e_{12}  \\
	e_{21} &1
\end{array}\right|
\la{tbt}
\ea
where we have defined
\be
e_{ij}=\exp\left[-\frac1{2\ka_N}(\br_i-\br_j)^2\right].
\ee
From Appendix \ref{aa}, one finds that
\ba
Z_2
&=& \frac1{2!}( z_1^2-z_2),
\la{z2}
\ea
where
\ba
z_n&=&\frac1{(2\sinh(nw/2))^D}=\left (\frac1{\e^{nw/2}-\e^{-nw/2}}\right )^D\nn\\
&=&\left (\frac{\e^{-nw/2}}{1-\e^{-nw}}\right )^D=\left (\frac{b^{n/2}}{1-b^n}\right )^D
\la{zn}
\ea
with $b=\exp(-w)$.

For $n=3$, one has
\ba
Z_3
&=&\frac1{3!}\frac1{(2\pi\ka_N)^{3D/2}}\int d\br_1d\br_2d\br_3\e^{-\mu_N(\br_1^2+\br_2^2+\br_3^2)}
\left|\begin{array}{ccc}
	1&e_{12} &e_{13} \\
	e_{21}&1 &e_{23}  \\
	e_{31}&e_{32} &1
\end{array}\right|\nn\\
&=&\frac1{3!}\frac1{(2\pi\ka_N)^{3D/2}}\int d\br_1d\br_2d\br_3\e^{-\mu_N(\br_1^2+\br_2^2+\br_3^2)}(1-3e_{12}e_{21}+2e_{12}e_{23}e_{31})
\ea
Again, from Appendix \ref{aa},
\ba
Z_3&=&\frac1{3!}(z_1^3-3z_2z_1+2z_3).
\la{z3}
\ea
Similarly one can show that for $n=4$,
\be
Z_4=
\frac1{4!}(z_1^4-6z_2z_1^2+3z_2^2 +8z_3z_1-6z_4).
\la{z4}
\ee

It has been known for sometime that the free fermion partition $Z_n$
in the {\it continuum} limit of $w\rightarrow \tau$ is given by\cite{for71}:
\be
Z_n=\frac1{n!}
\left|\begin{array}{ccccc}
	z_1&1  &0&0&\cdots \\
	z_2&z_1&2&0&\cdots \\	
	z_3&z_2 &z_1&3&\cdots \\	
	z_4&z_3 &z_2&z_1&\cdots \\
	\vdots&\vdots &\vdots&\vdots&\cdots \\		
\end{array}\right|
\la{detz}
\ee
and obeys the recursion\cite{for71,bor93}:
\ba
Z_n&=&\frac1{n}\sum_{i=1}^n(-1)^{i-1}z_iZ_{n-i}\la{rec}
\ea
with $Z_0=1$. By generalizing (\ref{tbt}) to the case of a $n\times n$ determinant, Chaudhary\cite{cha23}
has shown that the $n$-fermion {\it discrete} PIMC partition functions at {\it every} $N^{th}$ time-step, such as
(\ref{z2}), (\ref{z3}), (\ref{z4}), obey the same
recursion relation (\ref{rec}) and therefore also given by the same determinant form (\ref{detz}).
Since the discrete partition function (\ref{disz}) contains the continuum limit as its special case,
the above can be viewed as an alternative derivation of the fermion recursion relation directly 
from the path integral formalism. 

\section{Analytical thermodynamic and Hamiltonian energies}
\la{athe}

To compute the discrete PIMC thermodynamic energy at $\tau=N\ep$,
note that the universal coefficients $\ka(Nu)$ and $\mu(Nu)$ are functions of the
algorithm imaginary time
$w=Nu$.
Therefore,
one can write
\ba
E_n^T(\tau)&=&-\frac{\pa \log Z_n}{\pa\tau}=\frac{\pa w}{\pa \tau}\left(-\frac{\pa \log Z_n}{\pa w}\right),\nn\\
&=&\rho_T(\ep) E_n(w).
\la{eth}
\ea
where $\rho_T$ is the same prefactor for all $n$-fermion energy from (\ref{udef})
\be
\rho_T(\ep)=\frac{\pa w}{\pa \tau}=\frac{du}{d\ep}=\frac1{\sqrt{\za_1^2-1}}\frac{d\za_1}{d\ep}.
\la{dude}
\ee
Since the basic partition function $z_n$ (\ref{zn}) depends on $w$ only through $b=\e^{-w}$, 
one has the universal discrete $N^{th}$ time-step thermodynamic energy 
\ba
E_n(w)&=&-\frac{\pa \log Z_n(b)}{\pa b}\frac{\pa b}{\pa w}\nn\\
&=&b\frac{\pa \log Z_n(b)}{\pa b}
\la{ent}
\ea
for all short-time propagators. Each short-time propagator only modifies its argument $w$ via $u$.

For later comparisons, we will now only consider the case of $D=2$.
For $D=2$, the $n$-particle partition function is particularly simple:
\be
z_n(b)=\frac{b^{n}}{(1-b^n)^2}.
\ee
For one fermion, $\log Z_1=\log b-2\log (1-b)$ and (\ref{ent}) gives
\be
E_1(w)=(1+2\frac{b}{1-b})=\coth(w/2)=\frac{\sinh(w)}{\cosh(w)-1}.
\la{eng1}
\ee
For two fermions,
\ba
Z_2(w)&=&z_1^2(w)-z_2(w)=\frac{4 b^3}{(1-b)^4 (1+b)^2}
\la{az2}\\
E_2(w)&=&\frac{3+2b+3b^2}{1-b^2}=\frac{1+3\cosh(w)}{\sinh(w)}.
\la{eng2}
\ea
For three fermions
\ba
Z_3(w)&=&z_1^3-3z_1z_2+2z_3=\frac{6b^5(b^2+4b+1)}{(1-b)^6 (1+b)^2 (1+b+b^2)},\la{az3}\\
E_3(w)&=&\frac{5+31b+47b^2+50b^3+47b^4+31b^5+5b^6}{1+5b+5b^2-5b^4-5b^5-b^6}\nn\\
&=&\frac{25+47 \cosh(w)+31 \cosh(2 w)+5 \cosh(3 w)}{5 \sinh(w)+5 \sinh(2 w)+\sinh(3 w)}
\la{eng3}
\ea
For $n=4,5,6,10$, the lengthy expressions for the discrete energy are given in Appendix \ref{dneng}.

The fastest numerical method of computing the thermodynamic energy (\ref{eth}) at
imaginary time $\tau=N\ep$, corresponding to contracting (integrating over) $N$ 
short-time propagator $G_1(\ep)$, is to determine $u$ from (\ref{udef}),
evaluate the discrete energy $E_n(w)$ and multiply it by the prefactor $\rho_T(\tau/N)$.
However, in order to determine $E_n(w)$ analytically, one must also perform the subtractions in $Z_n(w)$,
as in (\ref{az2}) and (\ref{az3})
analytically, otherwise their direct numerical evaluations at $n>3$  would quickly exhaust machine-precision.

For $n$ and $N$ small, one can obtain analytical expressions for $E_n^T(\tau)$. 
Instead of (\ref{udef}), one observes that 
\be
\e^{u}=\za_1+\sqrt{\za_1^2-1}\quad{\rm and}\quad \e^{-u}=\za_1-\sqrt{\za_1^2-1}.
\la{expu}
\ee
For $n=2$, (\ref{eng2}) gives
\ba
E_2(Nu)
&=&\frac{2+3[\e^{Nu}+\e^{-Nu}]}{[\e^{Nu}-\e^{-Nu}]}\nn\\
&=&\frac{2+3[(\za_1+\sqrt{\za_1^2-1})^N+(\za_1-\sqrt{\za_1^2-1})^N]}
{[(\za_1+\sqrt{\za_1^2-1})^N-(\za_1-\sqrt{\za_1^2-1})^N]}.
\ea
For $N=2$,
\ba
E_2^T(2u)&=&E_2(2u)\frac{d\za_1}{d\ep}=
\frac{3\za_1^2-1}
{\za_1(\za_1^2-1)}\frac{d\za_1}{d\ep}.
\ea
The RHS are functions of $\ep$ through $\za_1(\ep)$.
For PA, $\za_1=1+\ep^2/2$ and $d\za_1/d\ep=\ep$, the two-bead ($N=2$) thermodynamic
energy in terms of $\tau=2\ep$ is then
\be
E_2^T(2u)=\frac{4}{\tau}\frac{(1+\frac38 \tau^2+\frac3{128}\tau^4)}
{(1+\frac3{16} \tau^2+\frac1{128}\tau^4)}.
\ee
Similarly, one can determine the 3- and 4-bead two-fermion energies:
\ba
E_2^T(3u)&=&\frac{4}{\tau}\frac{(1+\frac38 \tau^2+\frac1{36}\tau^4+\frac1{1944}\tau^6)}
{(1+\frac1{9} \tau^2)(1+\frac1{27} \tau^2)(1+\frac1{36} \tau^2)},\nn\\
E_2^T(4u)&=&\frac{4}{\tau}\frac{(1+\frac38 \tau^2+\frac{15}{512}\tau^4+\frac3{4096}\tau^6)+\frac3{524288}\tau^8)}
{(1+\frac1{32} \tau^2)(1+\frac1{64} \tau^2)(1+\frac1{8} \tau^2+\frac1{512}\tau^4)}.
\ea
Despite the appearance of exponentials in (\ref{eng2}), because of (\ref{expu}), all
discrete energies are just rational functions of $\tau$.

As shown in Ref.\onlinecite{chin23b}, for the 1D harmonic oscillator,
the Hamiltonian energy is given similarly as the thermodynamic energy (\ref{eth}), but with a different prefactor:
\ba
E_n^H(\tau)
&=&\rho_H(\ep) E_n(w).
\la{eh}
\ea
where
\be
\rho_H(\ep)=\frac12\left(\ga(\ep)+\frac1{\ga(\ep)}\right),
\ee
with $\ga(\ep)$ given by (\ref{gai}).

The ratio of the two energies at any discrete $N$ time step
is therefore given by 
\be
\frac{E_n^{H}(Nu)}{E_n^{T}(Nu)}=\frac{\rho_H(\ep)}{\rho_T(\ep)}
\ee
and is solely determined by the short-time propagator.

\begin{figure}[t]
	\includegraphics[width=0.49\linewidth]{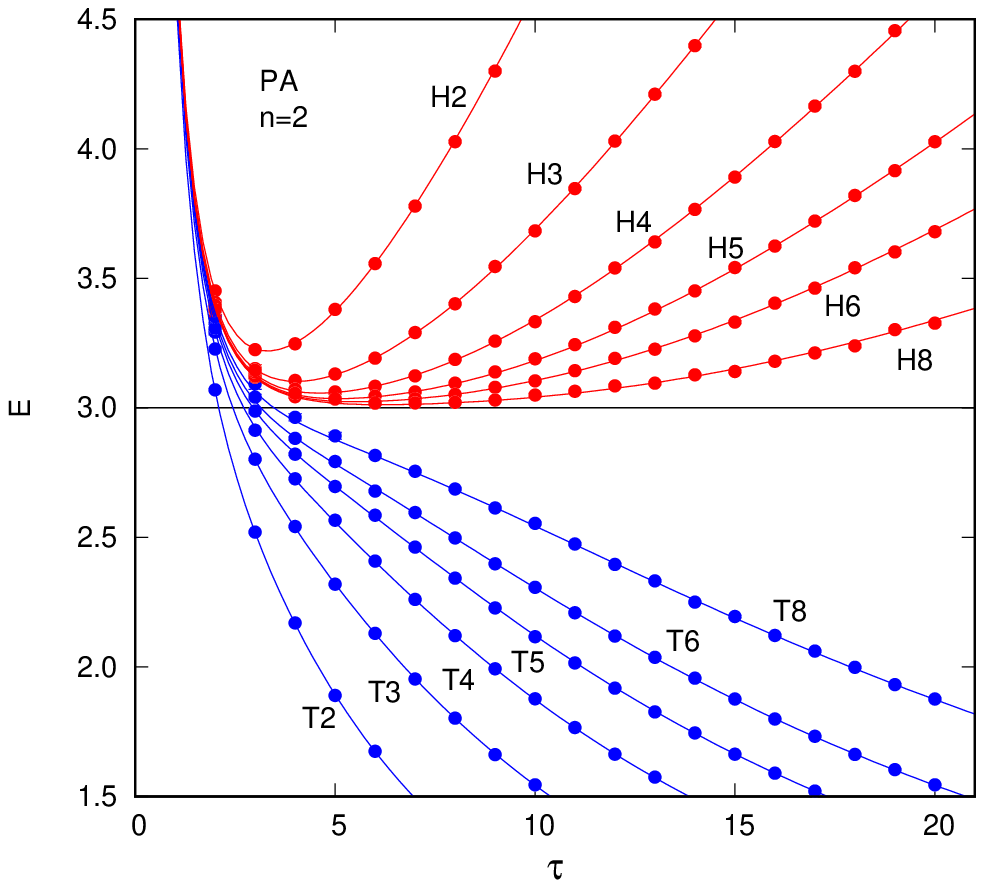}
	\includegraphics[width=0.49\linewidth]{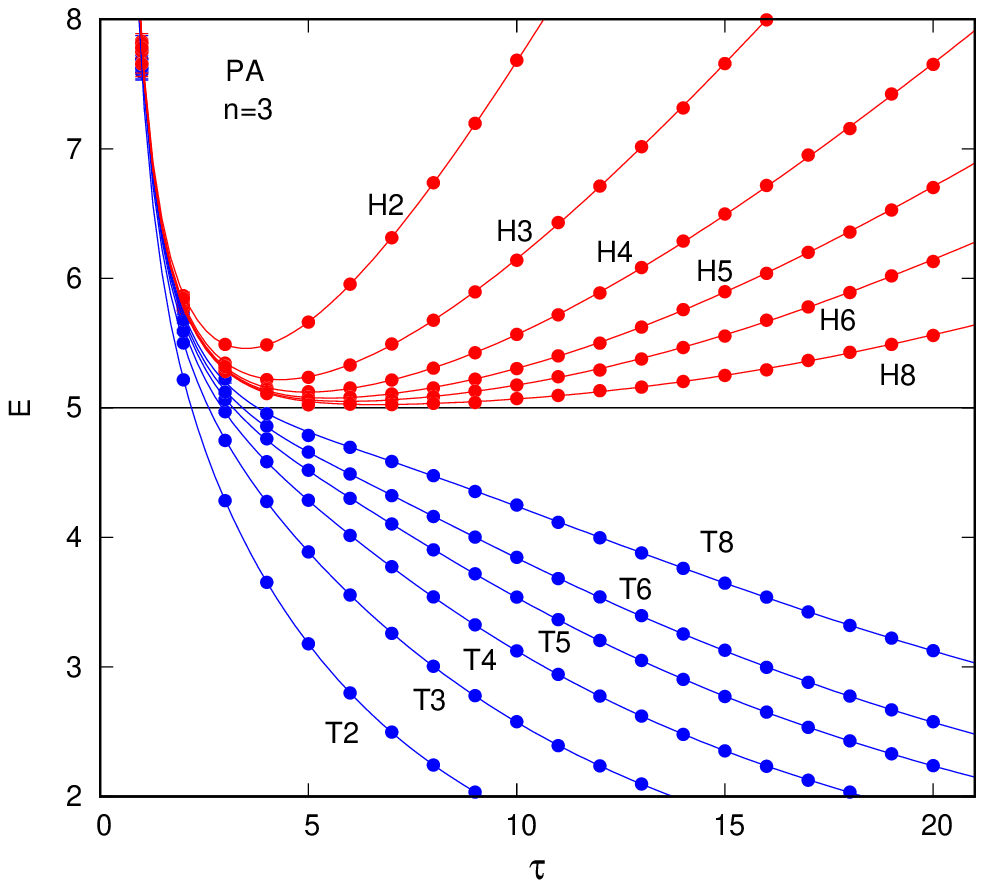}
	
	\includegraphics[width=0.49\linewidth]{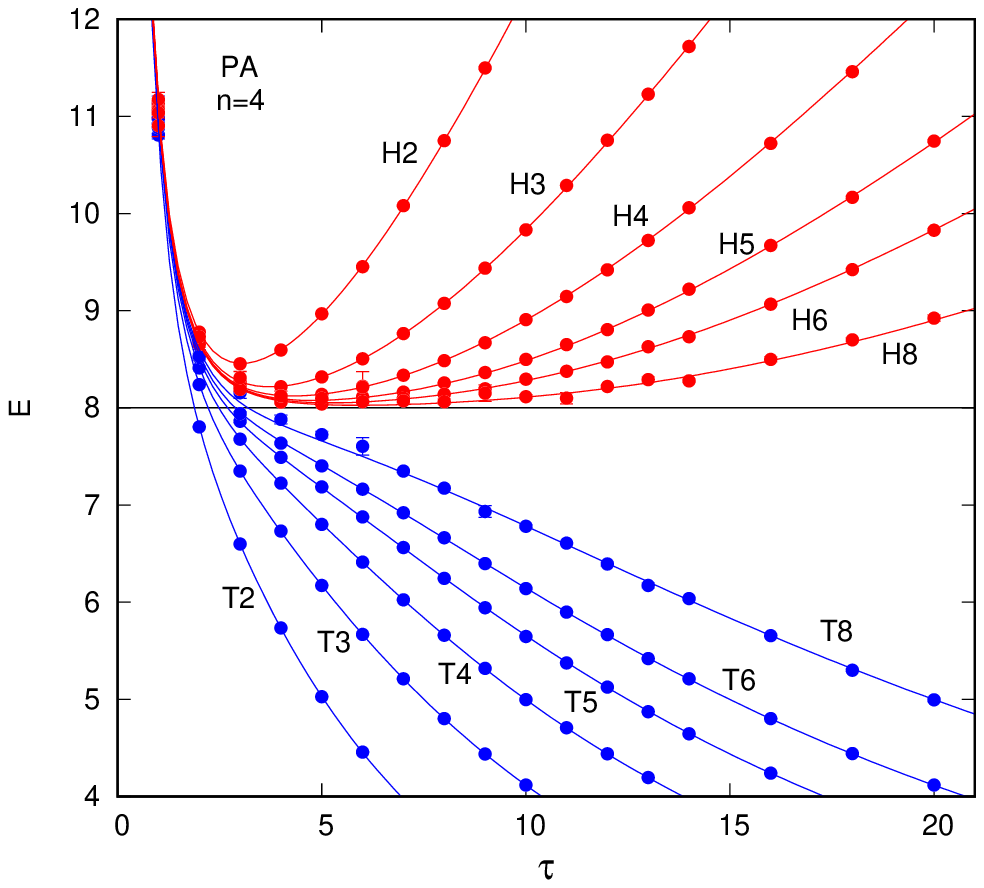}
	\includegraphics[width=0.49\linewidth]{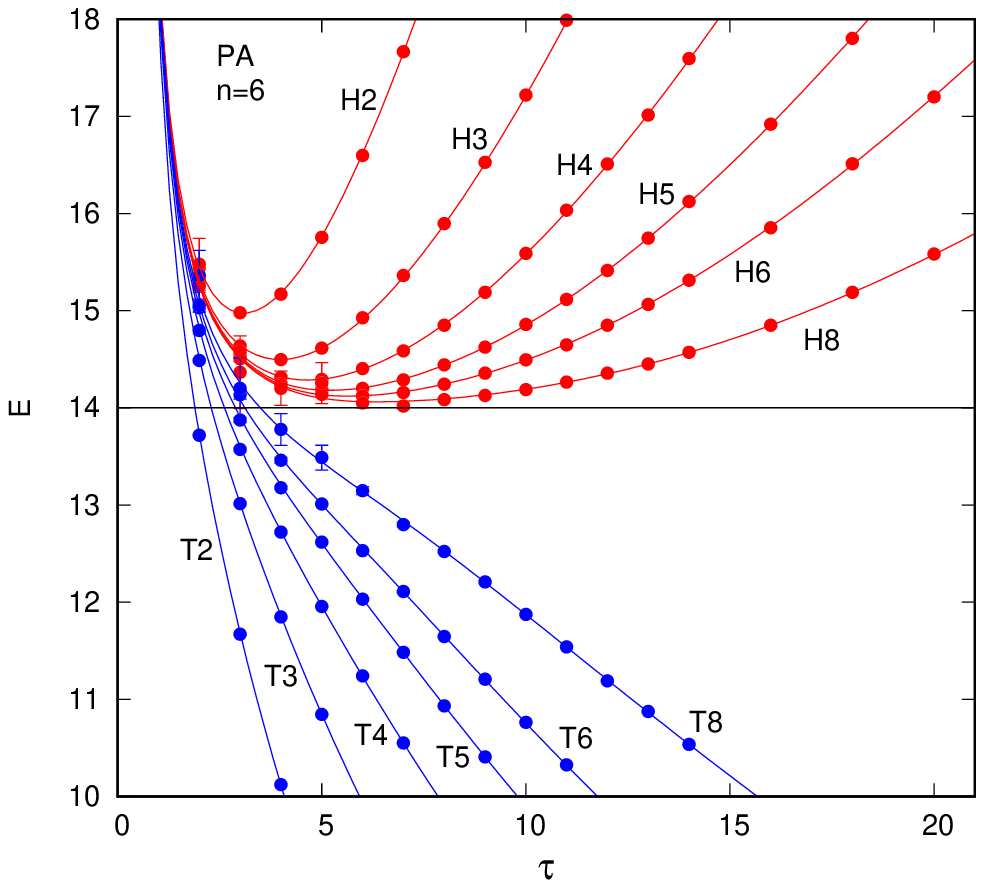}	
	\caption{ (color online) 
		The PA propagator's non-interacting 2, 3, 4 and 6-fermion $N$-bead thermodynamic 
		and Hamiltonian energies in a 2D harmonic oscillator, denoted by T$N$ and H$N$, 
		are plotted as a function of the imaginary time $\tau=N\ep$. 
		The data points are PIMC calculations and smooth curves are analytical results from (\ref{eth}) and (\ref{eh}).  
	}
	\la{engpa}
\end{figure}

For PA, this ratio is just $(1+\ep^2/8)$, giving the discrete Hamiltonian energies as
\be
E_n^H(Nu)=\left[ 1+\frac18 \left(\frac{\tau}{N}\right)^2\right ]E_n^T(Nu).
\la{ehh}
\ee
Since the harmonic oscillator is separable, the above is true for any number of
particles in any dimension, for all permutations, and therefore for fermions.

In Fig.\ref{engpa}, PA's thermodynamic and Hamiltonian energies 
for two, three, four and six fermions, up to $N=8$,
as determined above analytically, are compared with direct PIMC simulations. 
The agreements are perfect. 

Fig.\ref{engpa} also illustrates that, at these small $N$ values, PA's thermodynamic
energy showed no convergence whatsoever toward the exact ground
state fermion energies. By contrast, the energy minimum of the Hamiltonian energy is already
close to that of the ground state by $N=8$. As explained in Ref.\onlinecite{chin23b},
the thermodynamic energy converges to the same order as the short-time propagator, but
the Hamiltonian energy converges at twice the order of the algorithm and is
everywhere an upper bound to the ground state energy.

\begin{figure}[t]
	\includegraphics[width=0.49\linewidth]{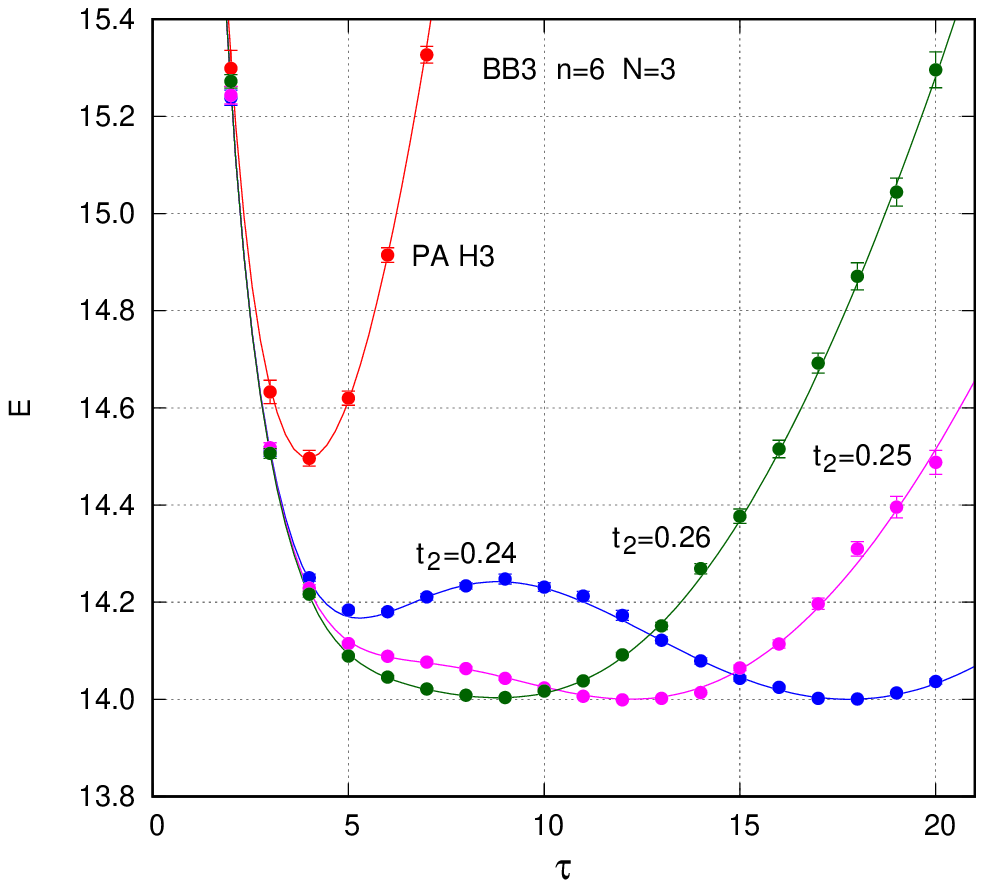}
	\includegraphics[width=0.49\linewidth]{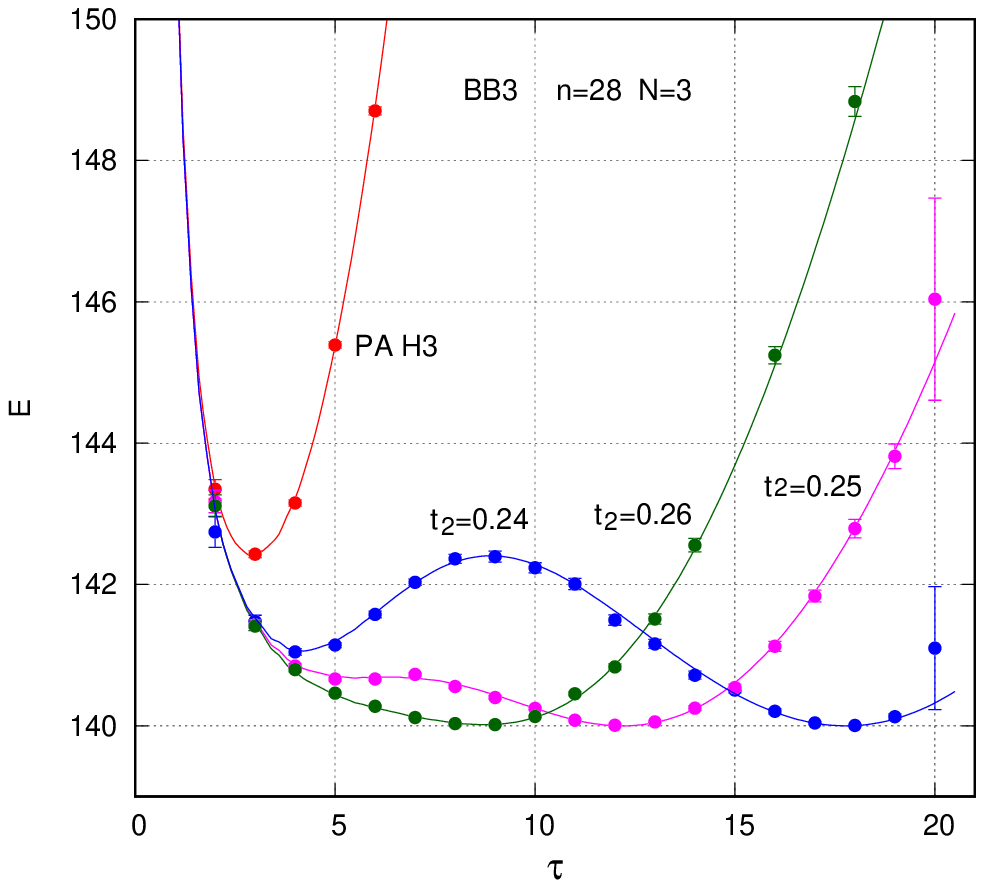}
	\caption{ (color online) 
		The Hamiltonian energies of the fourth-order Best 3-Bead (BB3) algorithm (see Appendix \ref{falg}) in solving for the ground state energy of 6 and 28 non-interacting fermions in a 2D harmonic oscillator. Notice that two energy minima are possible for $t_2=0.24$.  	}
	\la{n6opt}
\end{figure}

This means that if one uses fourth-order propagators, then the Hamiltonian energy will converge to
the eighth-order. Using the fourth-order algorithm BB3 as summarized in Appendix \ref{falg}, 
with only three beads (three free fermion propagators), one can obtain the exact ground state energy of up to 28 non-interacting fermions as shown in Fig.\ref{n6opt}. For $n=6$, since $E_6(w)$ is known analytically from Appendix \ref{dneng}, one can also compute its energy curves numerically, in also perfect agreement with PIMC data. 
For parameter values $t_1=0.26,0.25,0.24$, the minimum Hamiltonian energies can be determined
numerically to be
14.003,  14.0003 and 14.00002 respectively.
For $n=28$, given the complexity already shown by
$E_{10}(w)$ in Appendix \ref{dneng}, $E_{28}(w)$ is too lengthy to be derived analytically.
However, since $E_{28}(\tau)$ is just the energy of 28 non-interacting harmonic fermions, it can
be numerically computed via PIMC using the exact continuum propagator in only two beads, 
without any sign problem. Interpolate some 50 discrete values of $E_{28}(\tau)$ then produces the
smooth energy curves as shown in Fig.\ref{n6opt}. The sudden appearance of large error 
bars at $\tau=20$ is not due the sign problem nor limited Monte Carlo samplings. It is 
because for $n=28$ at such a large $\tau$, the determinant of the propagator matrix is 
too near zero and the matrix cannot be reliably inverted using double precision arithmetic.
This error plagues all large $n$ calculations at large $\tau$, but
would not appear if the code can be executed with greater arithmetic precision.

For non-interacting closed-shell fermions then, the ground state energy can be extracted 
by optimizing the Hamiltonian energy using this BB3 propagator, with seemingly no sign problem
for up to about 30 fermions.

\section{The fermion PIMC sign problem}
\la{sgpb}

Because the operator contraction is exact, there is no sign problem in the analytical calculations in the last section. All energies are just rational functions of $\tau$. In PIMC simulations, the partition function at $\tau=N\ep$
is a discrete path integral of $N$ short-time determinant propagators $G_1$ given by (\ref{shp}):
\be
Z=
\int d\x_1\cdots d\x_{N}\,
G_1(\x_1,\x_2;\dte)G_1(\x_2,\x_3;\dte)\cdots G_1(\x_N,\x_1;\dte).
\ee
Since the product of determinants is not necessary positive, one decomposes the integrand into its overall sign and
absolute value,
\ba
G_1(\x_1,\x_2;\dte)\cdots G_1(\x_N,\x_1;\dte)&=&\sgn\, |G_1(\x_1,\x_2;\dte)\cdots G_1(\x_N,\x_1;\dte)|\nn\\
&=& \sgn\, P(\x_1,\cdots \x_N).
\ea
Any observable ${\cal O}(\x)$ can then be computed as
\ba
\langle {\cal O}\rangle &=&\frac{ \int d\x_1\cdots d\x_{N}\bar{\cal O}\, \sgn\, P(\x_1,\cdots \x_N)}
{\int d\x_1\cdots d\x_{N} \sgn\, P(\x_1,\cdots \x_N)}\nn\\
 &=&\frac{ \int d\x_1\cdots d\x_{N}\bar{\cal O}\,\sgn\, P(\x_1,\cdots \x_N)/\int d\x_1\cdots d\x_{N} P(\x_1,\cdots \x_N)}
{\int d\x_1\cdots d\x_{N} \sgn\, P(\x_1,\cdots \x_N)/\int d\x_1\cdots d\x_{N} P(\x_1,\cdots \x_N)}\nn\\
&=&\frac{\langle \bar{\cal O}\, \sgn\rangle_P}{\langle \sgn\rangle_P},
\la{sgnavg}
\ea
where $\bar{\cal O}$ is the operator averaged over each short-time propagator (the bead-average) 
\be
\bar{\cal O}=\frac1{N}\sum_{i=1}^N{\cal O}(\x_i).
\ee
Both the numerator and denominator of (\ref{sgnavg}) can be computed by the Monte Carlo 
method because $P(\x_1,\cdots \x_N)$ is non-negative. The sign problem arises only when
the average sign $s=\langle \sgn\rangle_P\rightarrow 0$ such that (\ref{sgnavg})
can no longer be evaluated with stability. 

\begin{figure}[htb]
	\includegraphics[width=0.49\linewidth]{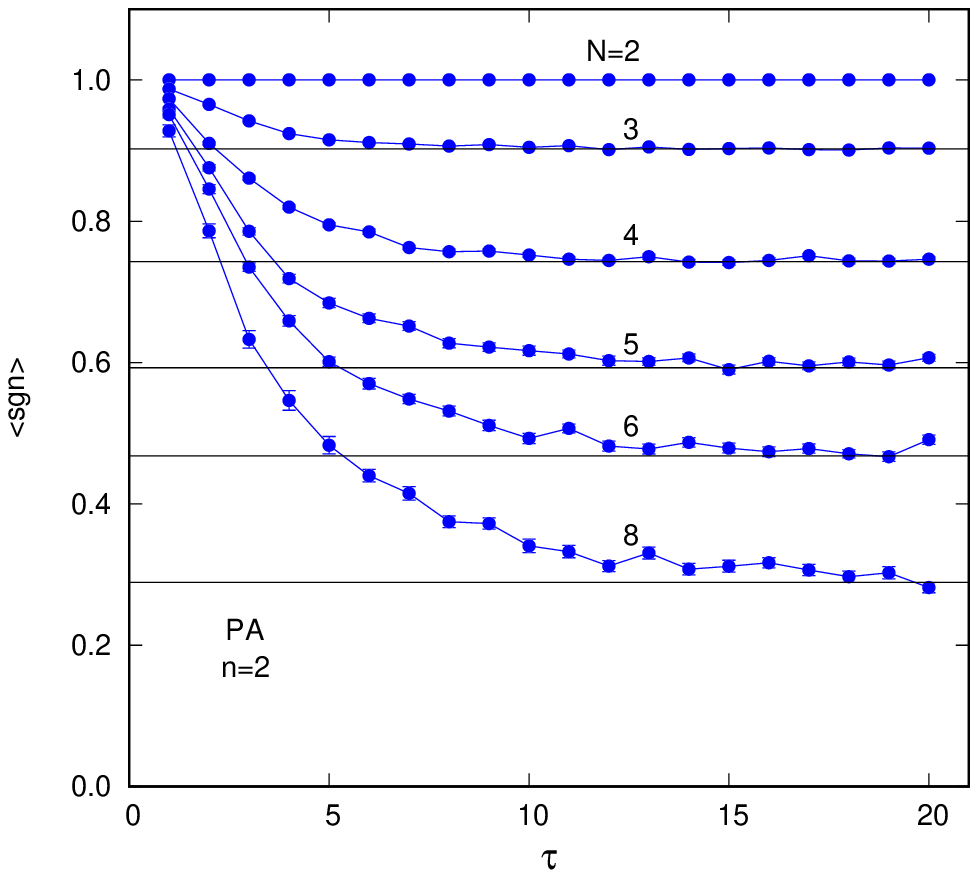}
	\includegraphics[width=0.49\linewidth]{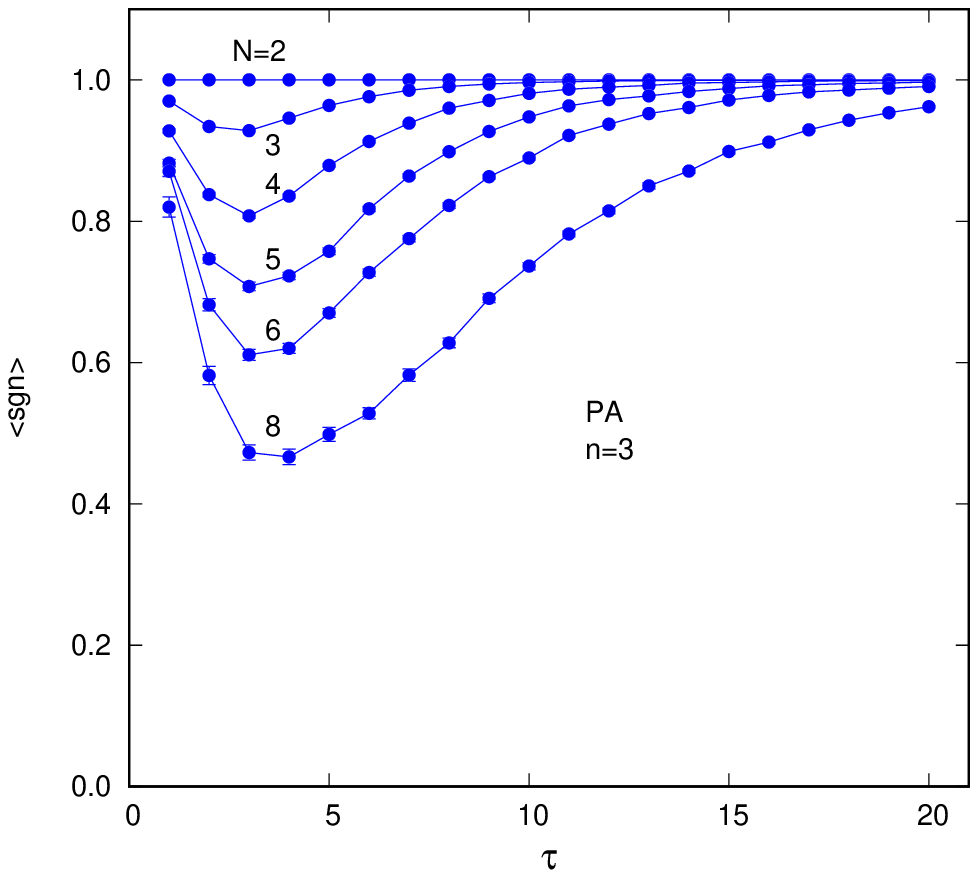}
	
	\includegraphics[width=0.49\linewidth]{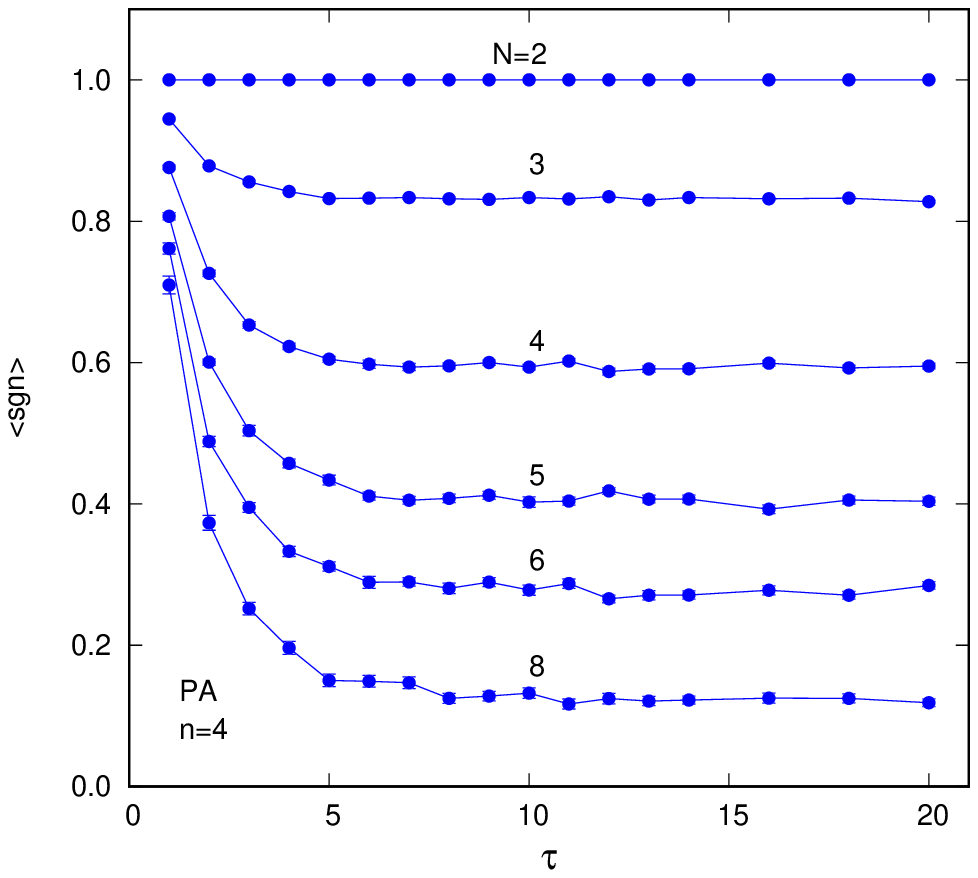}
	\includegraphics[width=0.49\linewidth]{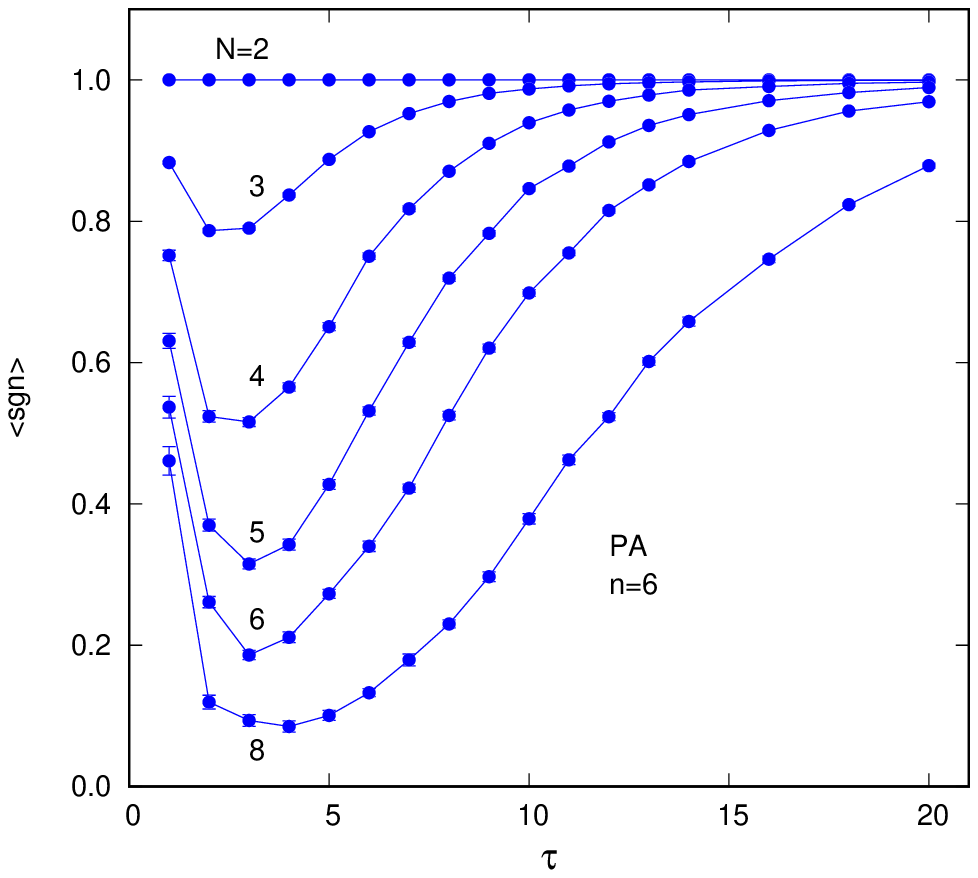}
	\caption{ (color online) 
		The PA propagator's average sign $\langle \sgn\rangle$ for $n=2,3,4,6$ non-interacting 
		harmonic fermions at various bead number $N$
		as a function of $\tau$. The horizontal black lines in the two-fermion case are average 
		sign values at large $\tau$ deduced in Appendix \ref{ass}.
	}
	\la{ttsgn}
\end{figure}

The average sign values for the 2, 3, 4 and 6-fermion calculation in Fig.\ref{engpa} are shown on
Fig.\ref{ttsgn}. For up to $N=8$, the average sign is nowhere near zero. This is why the energy calculations in
Fig.\ref{engpa} (and Fig.\ref{n6opt}) are without problem. However, there are at least two surprises in
Fig.\ref{ttsgn}: 1) It is generally expected that $s$ should deceases exponentially\cite{tro05,dor19} 
with increasing $n$ and $N$.
In the 2 and 4-fermion cases, it was unexpected to find that $s$ actually levels off to a constant
value at large $\tau$. The predicted $s$ values for the two-fermion case are derived in Appendix \ref{ass}. 
This was not noticed before because most calculations\cite{dor19} were done at $\tau<<10$. 
2) More surprisingly, for $n=3$ and 6, the average sign does not decrease monotonically, but reaches 
a sign minimum then reverts back to 1 at large $\tau$. Since $n=3$ and 6 are a closed-shell harmonic 
states in 2D, this immediately suggests that one should check whether this is true for all closed-shell
fermion states. 

\begin{figure}[htb]
	\includegraphics[width=0.49\linewidth]{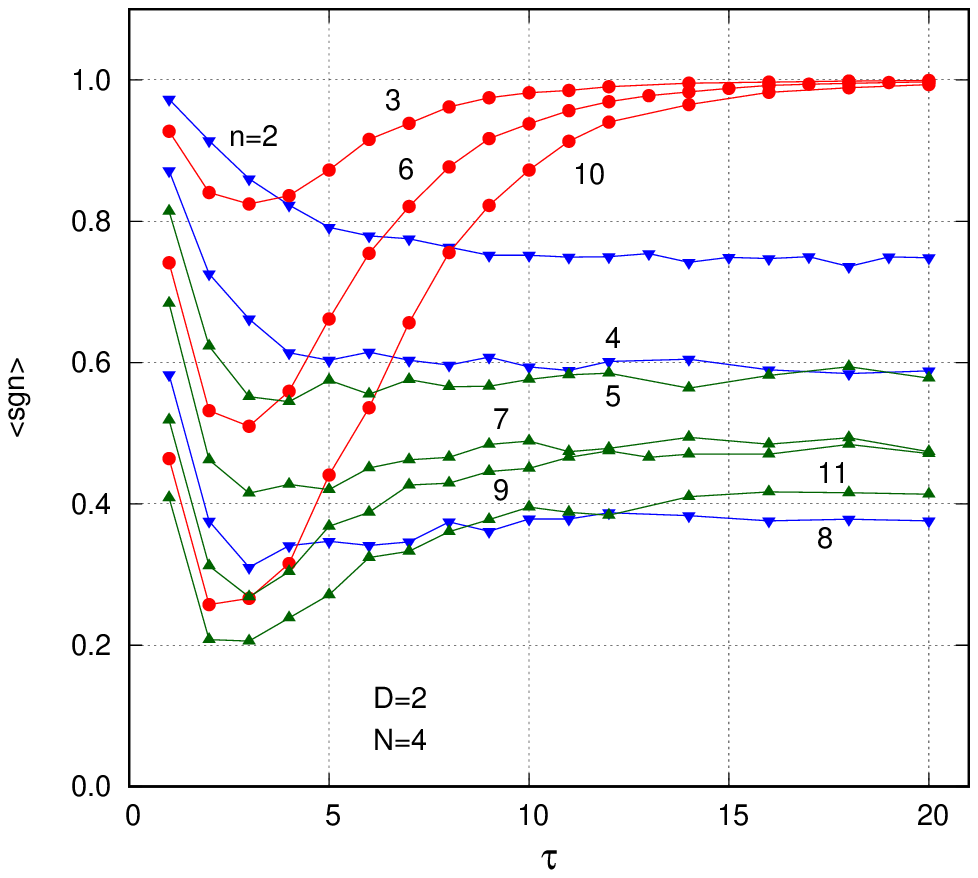}
	\includegraphics[width=0.49\linewidth]{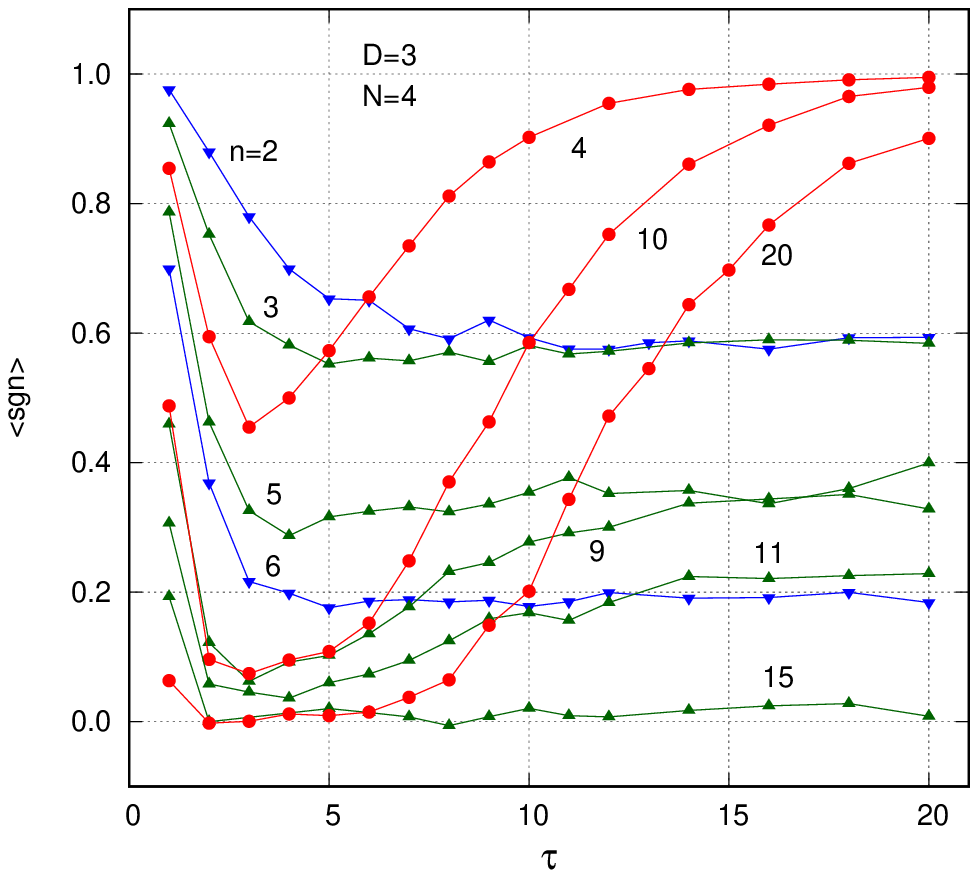}
	\caption{ (color online) 
		The average sign of a $N=4$ bead calculation in 2D
		and 3D for various $n$ non-interacting harmonic fermions.
		The average sign reverts back to one for closed-shell number
		of fermions in the large $\tau$ limit.}
	\la{sign23}
\end{figure}

This is indeed the case for both 2D and 3D as shown in Fig.\ref{sign23}.
One observes that at $\tau=1$, in both 2D and 3D, $s$ decreases with increasing $n$,
which is the conventional understanding\cite{tro05,dor19}. However, Fig.\ref{sign23} clearly shows
that this is not the case for closed-shell states at large $\tau$.

\section{Why no sign problem for closed-shell states?}
\la{nosgn}

The surprising reason why there is no sign problem in closed-shell states,
is the same reason why there is no sign problem for fermions in one dimension\cite{chin24}.
As will be shown below, the reason why fermion in one dimension has no sign problem is
because the sign of the propagator is given
by a {\it single} number, and not by a {\it sum} of many numbers (a dot-product).
Similarly, in more than one dimension, the free fermion propagator's sign in the first closed-shell state 
is also given by a single number, not a sum of numbers.

Let's begins with the two-fermion free propagator in dimension $D$,
which gives the clearest insight into the nature of the sign problem.
From (\ref{gdet}), one has
\ba
G_0(\br_1^\prime,\br_2^\prime,\br_1,\br_2;\dte)
&=&\frac12\frac1{(2\pi\dte)^2}
\det\left(\begin{array}{cc}
	\e^{-(\br'_1-\br_1)^2/(2\dte)} &  \e^{-(\br'_1-\br_2)^2/(2\dte)}\\
	\e^{-(\br'_2-\br_1)^2/(2\dte)} & \e^{-(\br'_2-\br_2)^2/(2\dte)}
\end{array}\right),\la{twofd}\\
&=&\frac12\frac1{(2\pi\dte)^2}\e^{-\frac1{2\dte}\left[(\br_1^\prime-\br_1)^2
	+(\br_2^\prime-\br_2)^2\right] }
\left(1-\e^{-\frac1{\dte}\br_{21}^\prime\cdot\br_{21} } \right),
\la{propr2}
\ea
where $\br_{21}'=\br_2'-\br_1'$ and $\br_{21}=\br_2-\br_1$. Its sign is
given by the sign of the round bracket above, 
which in turn is
given by the sign of $\br_{21}'\cdot\br_{21}$. Therefore one has
\be
\sgn\Bigl(G_0(\br_1',\br_2',\br_1,\br_2;\dte)\Bigr)
=\sgn(\br_{21}'\cdot\br_{21}),
\ee
independent of $\ep$.
For a closed loop of three propagators,
\ba
\sgn\Bigl(G_0(\br_{21},\br_{21}')G_0(\br_{21}',\br_{21}'') G_0(\br_{21}'',\br_{21})\Bigr)
&=&\sgn\Bigl((\br_{21}\cdot\br_{21}')(\br_{21}'\cdot\br_{21}'')(\br_{21}''\cdot\br_{21})\Bigr),\nn\\
&=&|\br_{21}|^2|\br_{21}'|^2|\br_{21}''|^2\sgn\Bigl(\cos\theta\cos\theta'\cos\theta''\Bigr),
\la{tsign}
\ea
one has the sign problem because the cosine functions resulting from the dot products can have both signs.
This is true for any $D$-dimension vectors $\br_{21}, \br_{21}'$, etc., except for $D=1$.  
In one dimension, there is no dot product, no cosine functions, 
\ba
\sgn\Bigl(G_0(x_{21},x_{21}')G_0(x_{21}',x_{21}'') G_0(x_{21}'',x_{21})\Bigr)
&=&\sgn\Bigl((x_{21}x_{21}')(x_{21}'x_{21}'')(x_{21}'' x_{21})\Bigr),\nn\\
&=&\sgn\Bigl((x_{21})^2(x_{21}')^2(x_{21}'')^2\Bigr)\ge 0,
\la{loops}
\ea
and therefore no sign problem. Note that $x_{ij}$ can be viewed as a {\it signed-distance}.

There is no sign problem with (\ref{loops}) because the sign is solely determined by a {\it single} number,
a product of both relative distances.
When propagators are looped, that single number is paired between adjacent relative distances and squared, 
which is then non-negative.
There is a sign problem in (\ref{tsign}) because the sign is determine by a {\it sum} of numbers,
a dot product of two relative vectors. When propagators are looped, the dot product
is always between adjacent relative vectors, resulting in a product of cosine functions not necessarily positive.

Based on this insight, one can prove in Appendix \ref{signp}, 
that in $D$-dimension, the first closed-shell state with $n=D+1$ fermions 
has no sign problem in the large $\tau$ limit.
The sign of the propagator is given by the product of
two {\it signed hyper-volumes} formed by two sets of $D$-dimension relative vectors. 
The sign of a loop of propagators is then the sign
of a product of hyper-volume squared, which is then non-negative, exactly similar 
to the one dimensional case. For {\it higher} closed-shell states, while no proof currently exists, 
Fig.\ref{sign23} provided direct numerical evidence that this is also the case.

\section{The sign problem with interaction}
\la{sgint}

To see how interactions alter the sign problem, consider first the case of pairwise harmonic interaction.
The use of the exact {\it continuum} path integral to study fermions with pairwise 
harmonic potential has been studied extensively by
Brosens, Devreese and Lemmens\cite{bro97,bro98}. This work focuses instead, on understanding 
the sign problem at each {\it discrete} time step of PIMC. For completeness, we give below a 
succinct summary of the needed results.

The Fortran code for this work is written for a $n$-particle Hamiltonian
with a general pairwise interaction $\lm |\br_i-\br_j|^k$, so that by simply setting  $k=2,-1,-3$, the same PIMC code would generate results for the harmonic, Coulomb and dipolar interaction respectively. This means that, once the harmonic force case agrees with analytical results derived below, one can then be assured that the code is working properly for the other interactions.
 
For the harmonic pairwise interaction, one can write the Hamiltonian as 
\be
H=T+V=\frac1{2m}\sum_{i=1}^n\p_i^2+\frac12 \sum_{i=1}^n{\br_i}^2+\frac12 \lm\sum_{i=1}^n\sum_{j=1}^n(\br_i-\br_j)^2
\ee
with $m=1$.
Let $\rr$ be the center-of-mass (CM) vector and $\bu_i$ the position relative to $\rr$:
\be
\rr=\frac1{n}\sum_{i=1}^n\br_i,\qquad\quad \bu_i=\br_i-\rr.
\ee
It then follows that $\sum_{i=1}^n \bu_i=0$ and 
\ba
V&=&\frac12 \sum_{i=1}^n(\rr+\bu_i)^2+\frac12 \lm\sum_{i=1}^n\sum_{j=1}^n(\bu_i-\bu_j)^2\nn\\
&=&\frac12 (\sqrt{n} \rr)^2+\frac12(1+ 2n\lm)\sum_{i=1}^n \bu_i^2.
\ea
Similarly, the CM momentum is ($M=n$)
\be
\pp=M\dot\rr=M\frac1{n}\sum_{i=1}^n\dot \br_i=\sum_{i=1}^n\p_i,
\ee
and the relative to CM momentum $\s_i$
\ba
\s_i= m\dot \bu_i=m\dot \br_i-m\dot \rr=\p_i-\frac{1}{n}\pp
\ea
also satisfies
\be
\sum_{i=1}^n \s_i=0,
\ee
giving the kinetic energy as
\ba
T&=&\frac1{2} \sum_{i=1}^n(\frac1{n}\pp+\s_i)^2
=\frac1{2n} \pp^2+\frac1{2}\sum_{i=1}^n \s_i^2.
\ea
The total Hamiltonian is then
\be
H=\frac1{2n} \pp^2+\frac12 n \rr^2+\frac1{2}\sum_{i=1}^n \s_i^2+\frac12(1+ 2n\lm)\sum_{i=1}^n \bu_i^2.
\ee
Comparing the center-of-mass Hamiltonian to
$
\frac1{2M}\pp^2+\frac12 M\Omega^2 \rr^2,
$
one sees that $M=n$ and $\Omega=1$. Therefore the
center-of-mass energy is $\Omega=1$ and the relative energy 
\be
\w=\sqrt{1+2n\lm}
\la{omg}
\ee
is {\it $n$-dependent}. For attractive interactions, $\lm>0$ and $\w>1$.
For repulsive interactions, $-1/(2n)<\lm <0$ and $0<\w<1$,
which means that the pairwise repulsion cannot be too strong, otherwise
the repulsion will blow up the harmonic confinement. For the ease of knowing the
exact ground state energy, we will generally fix $\w$ for all $n$ and set $\lm=(\w^2-1)/(2n)$.

For the case where each component potential is generalized to
\be
V=\frac12 x^2\rightarrow\frac12\w^2x^2
\ee
it is only necessary to replace
\be
\mu_N\rightarrow\mu_N^*= \w^2\mu_N,
\ee
in all relevant equations in Sect.\ref{upp}:
\ba
\za_N^*&=&1+\ka_N\mu_N^*=1+\ka_N\mu_N\w^2,\nn\\
u^*&=&\ln\left(\za^*_1(\ep)+\sqrt{{\za^*_1}^2(\ep)-1}\right),\la{ustr}\\
\ga^*&=&\frac{\sqrt{ {\za^*_1}^2-1}}{\ka_1}.
\la{gstr}
\ea
For $n$ fermions, one fills the relative energy spectrum $(n_x+n_y+1)\w$ in succession
but with its lowest energy state replaced by
the center-of-mass energy\cite{bro98}.
The thermodynamic energy is then given by
\ba
E_n^T(\tau)
&=& \frac1{\sqrt{\za_1^2-1}}\frac{d\za_1}{d\ep} E_1(Nu)+\frac1{\sqrt{{\za^*_1}^2-1}}\frac{d\za^*_1}{d\ep}\Bigl(E_n(Nu^*)-E_1(Nu^*)\Bigr)
\ea
For PA $\za_1=1+\ep^2/2$, $\za^*_1=1+\ep^2\w^2/2$, the above reduces to 
\ba
E_n^T(\tau)
&=& \frac{1}{\sqrt{1+\ep^2/4}} E_1(Nu)+\frac{\w}{\sqrt{1+\w^2\ep^2/4}}\Bigl(E_n(Nu^*)-E_1(Nu^*)\Bigr).
\la{thw}
\ea
Note that since $1/\sqrt{1+x}=1-x/2+\cdots$, the convergence of the thermodynamic energy is second-order in $\ep$ and
from below the exact energy.

\begin{figure}[b]
	\includegraphics[width=0.49\linewidth]{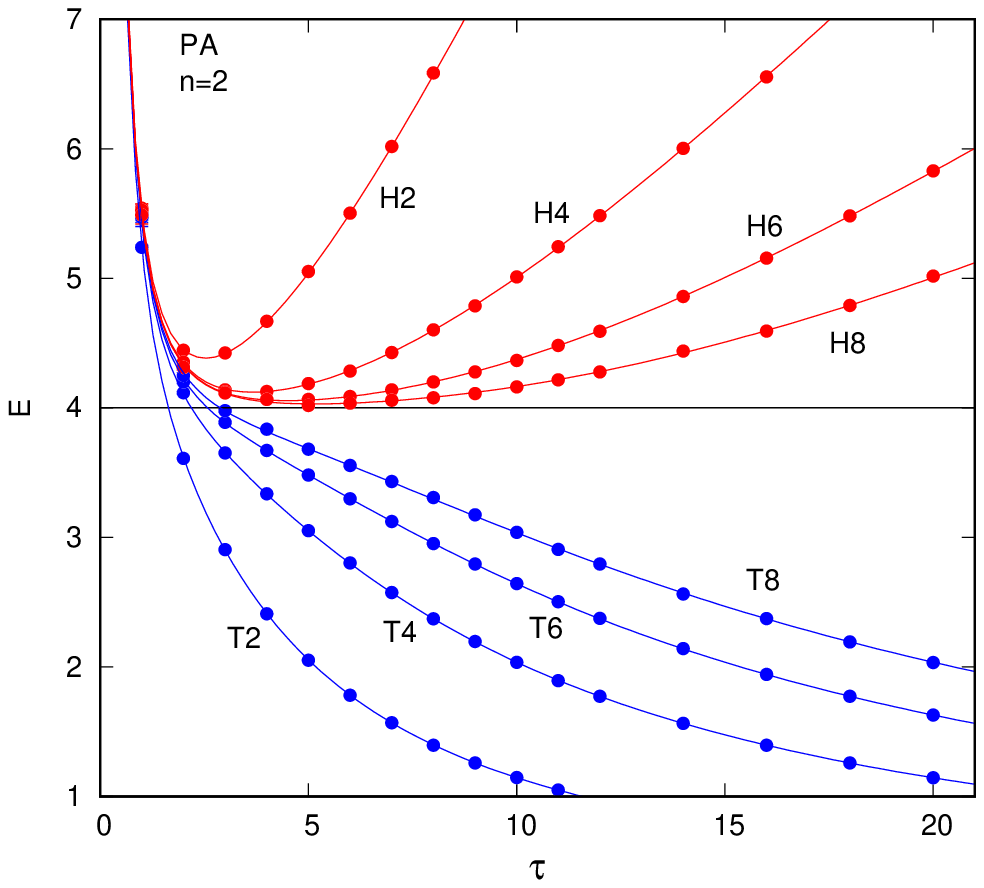}
	\includegraphics[width=0.49\linewidth]{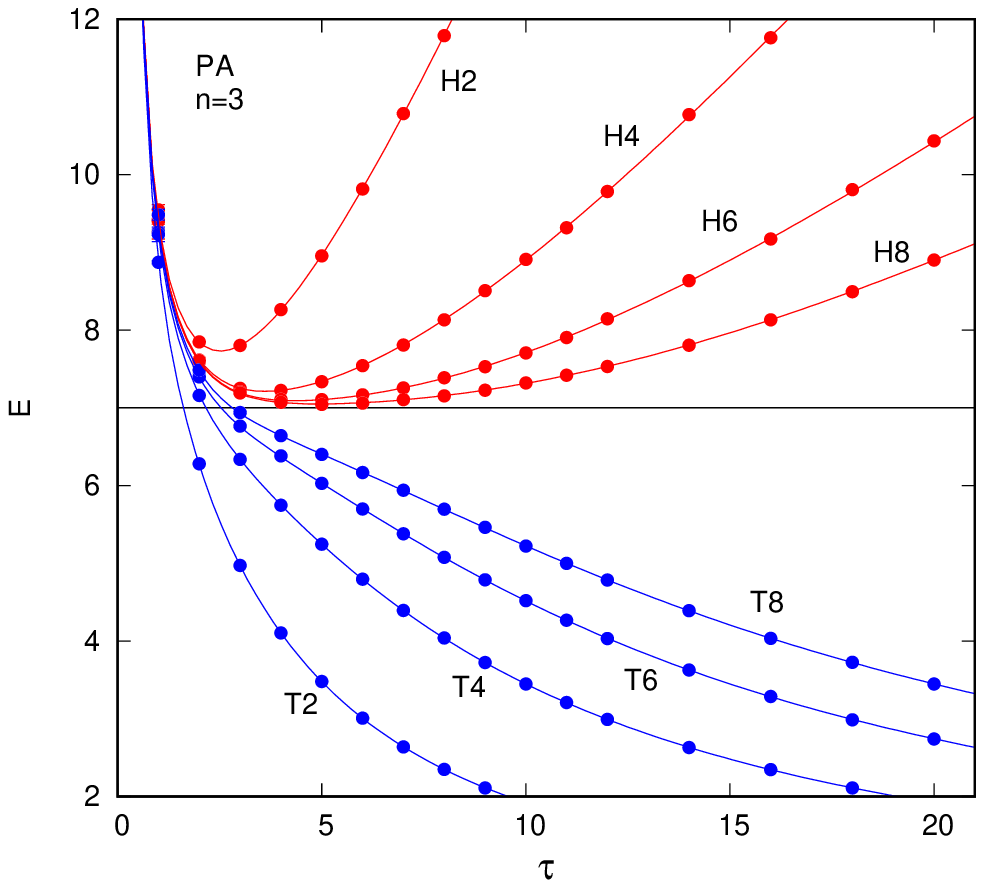}
	
	\includegraphics[width=0.49\linewidth]{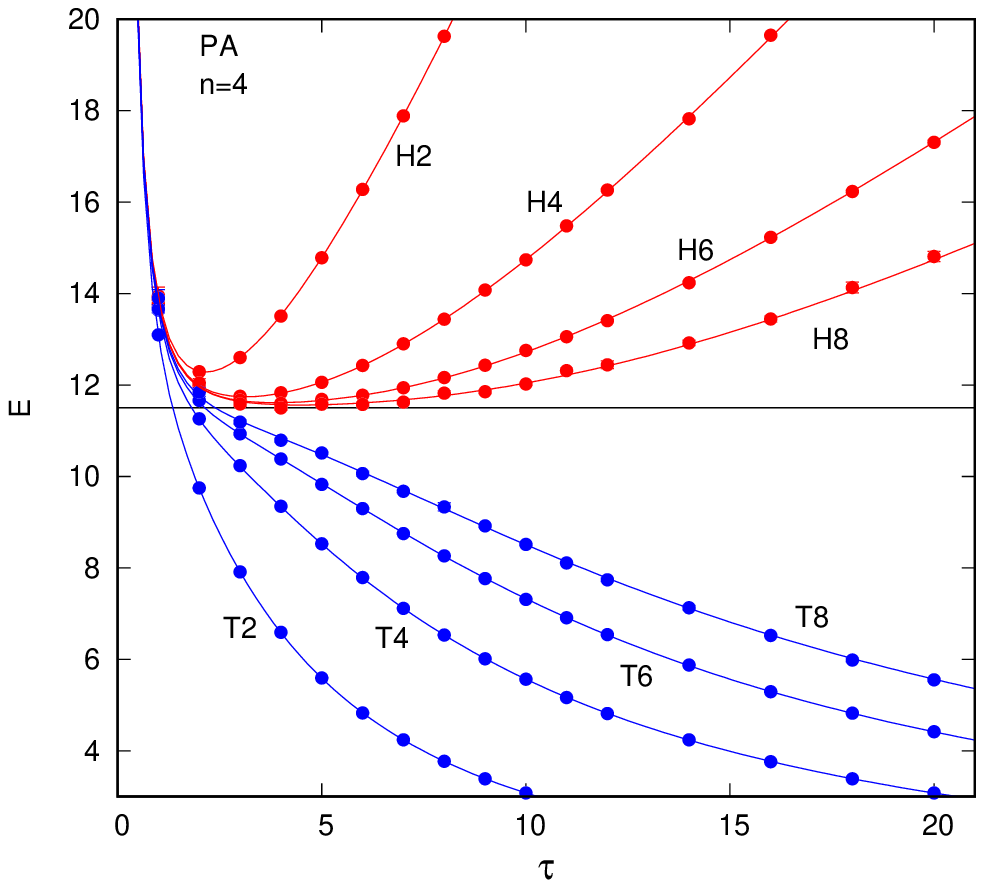}
	\includegraphics[width=0.49\linewidth]{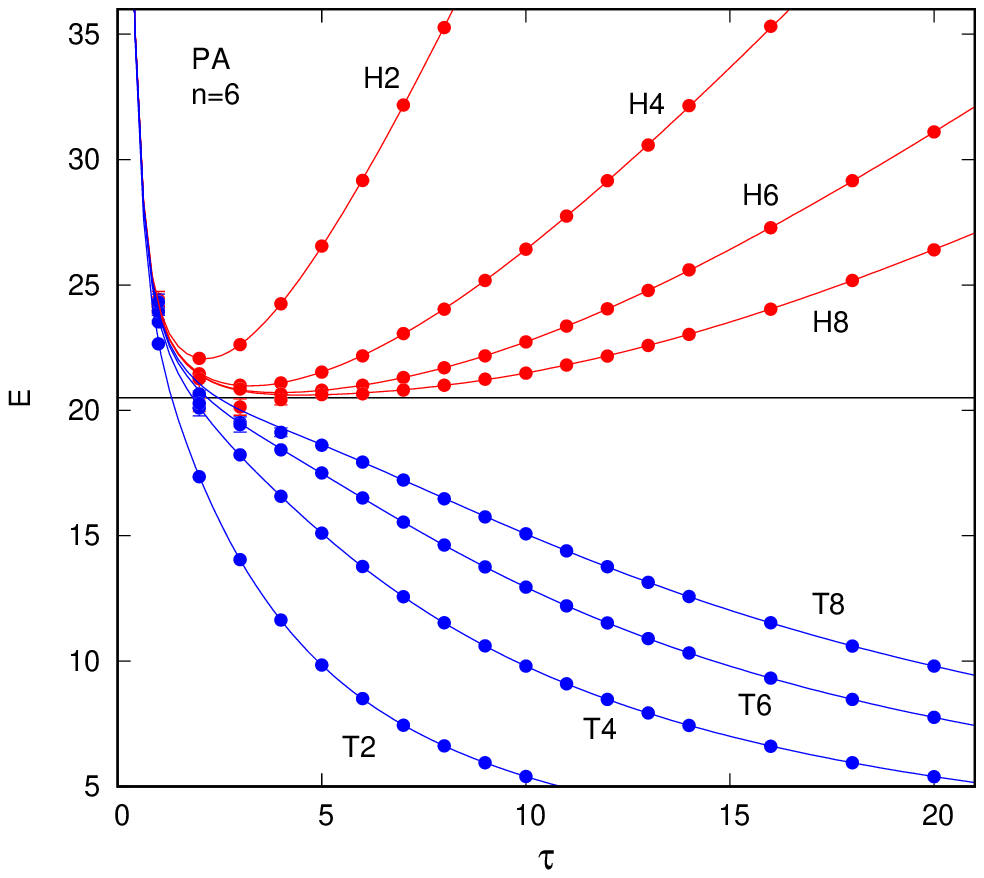}	
	\caption{ (color online) 
		The PA propagator's $n=$2, 3, 4 and 6-fermion $N$-bead thermodynamic 
		and Hamiltonian energies in a 2D harmonic oscillator with pairwise potential $\lm(\br_i-\br_j)^2$.
		The attactive potential strength is set to $\lm=5/(8n)$ so that $\w=3/2$ for all $n$.
		The data points are PIMC calculations and smooth curves are analytical results from 
		(\ref{thw}) and (\ref{hw}).  
	}
	\la{enghar}
\end{figure}

In the continuum limit of $\ep\rightarrow 0$ and $\tau\rightarrow\infty$, one has from (\ref{eng1}),
(\ref{eng2}), (\ref{eng3}), (\ref{engn}),
$
E_1(\tau)\rightarrow 1,
E_2(\tau)\rightarrow 3,
E_3(\tau)\rightarrow 5,
E_4(\tau)\rightarrow 8,
E_6(\tau)\rightarrow 14.
$
The above (\ref{thw}) then gives the exact ground state fermion energies according to the degeneracy of the 2D harmonic oscillator:
\be
E_2^T\rightarrow 1+2\w,\quad
E_3^T\rightarrow 1+4\w,\quad
E_4^T\rightarrow 1+7\w,\quad
E_6^T\rightarrow 1+13\w.
\la{engs}
\ee

A careful re-deriving of the Hamiltonian energy as in Ref.\onlinecite{chin23b} with $\w\neq 1$ 
gives
\ba
E_n^H(\tau)
&=&  \frac12(\ga+\frac1{\ga}) E_1(Nu)+\frac{\w}2(\frac{\ga^*}{\w}+\frac{\w}{\ga^*})\Bigl(E_n(Nu^*)-E_1(Nu^*)\Bigr).
\la{hhw}
\ea
For PA, one has explicitly
\ba
E_n^H(\tau)
&&=\frac12\left(\sqrt{1+\ep^2/4}+\frac{1}{\sqrt{1+\ep^2/4}}\right) E_1(Nu)\nn\\
&&\qquad+\frac{\w}2\left(\sqrt{1+\ep^2\w^2/4}+\frac{1}{\sqrt{1+\ep^2\w^2/4}}\right)\Bigl(E_n(Nu^*)-E_1(Nu^*)\Bigr).
\la{hw}
\ea
Since $(1+x)^{1/2}+(1+x)^{-1/2}=1+ x^2/16+\cdots$, the convergence is now fourth-order in $\ep$ and from
above the exact energy.

\begin{figure}[b]
	\includegraphics[width=0.49\linewidth]{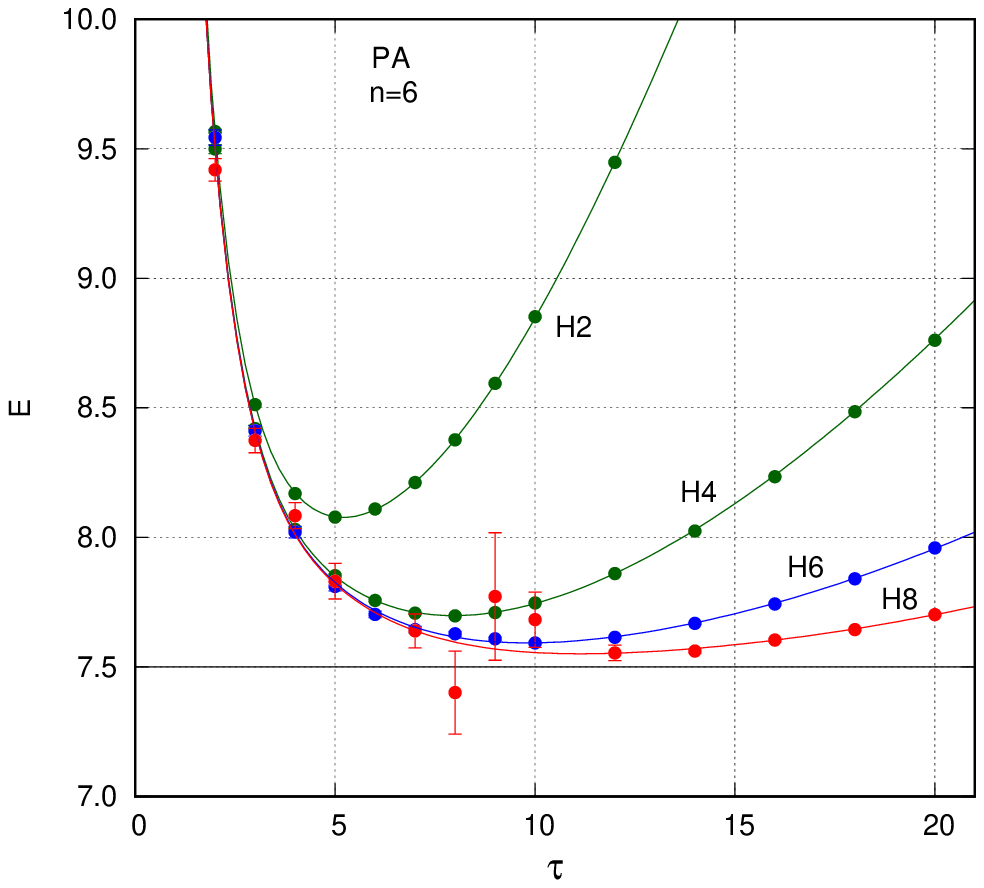}
	\includegraphics[width=0.49\linewidth]{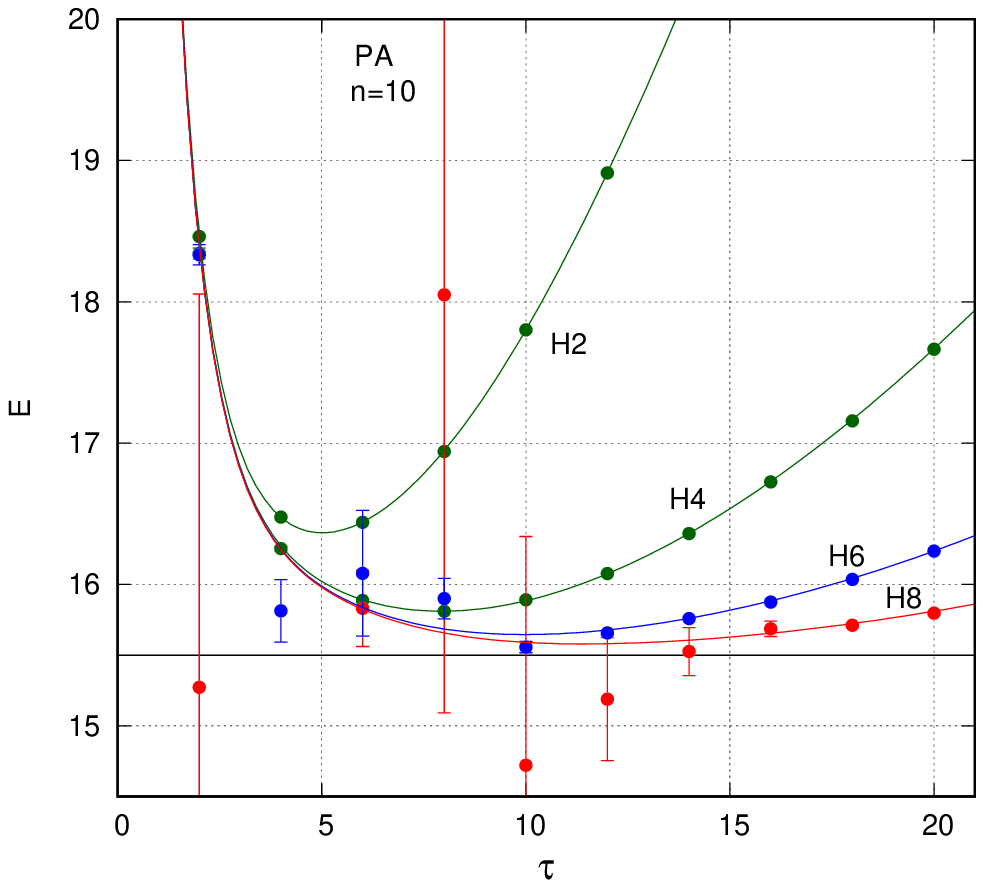}
	\caption{ (color online) 
		The $n=6$ and 10-fermion $N$-bead PA
		Hamiltonian energies in a 2D harmonic oscillator with repulsive pairwise potential $\lm(\br_i-\br_j)^2$.
		The repulsive potential strength is set to $\lm=-3/(8n)$ so that $\w=1/2$ for all $n$. 
		The data points are PIMC calculations and smooth curves are analytical results given by (\ref{hw}).
	}
	\la{neg610pa}
\end{figure}

The energy convergences of these four $n$ cases are shown in Fig.\ref{enghar},
with $\lm=5/(8n)$ so that $\w=3/2$ for all $n$ values. 
The exact ground state energies according to (\ref{engs}),
4, 7, 11.5 and 20.5, are shown as horizontal black lines in Fig.\ref{enghar}.
Again the PIMC results are in perfect agreement with
analytical thermodynamic and Hamiltonian energies given by
(\ref{thw}) and (\ref{hw}) respectively. As shown below, 
the sign problem in these cases are not very severe. 

\begin{figure}[b]
	\includegraphics[width=0.49\linewidth]{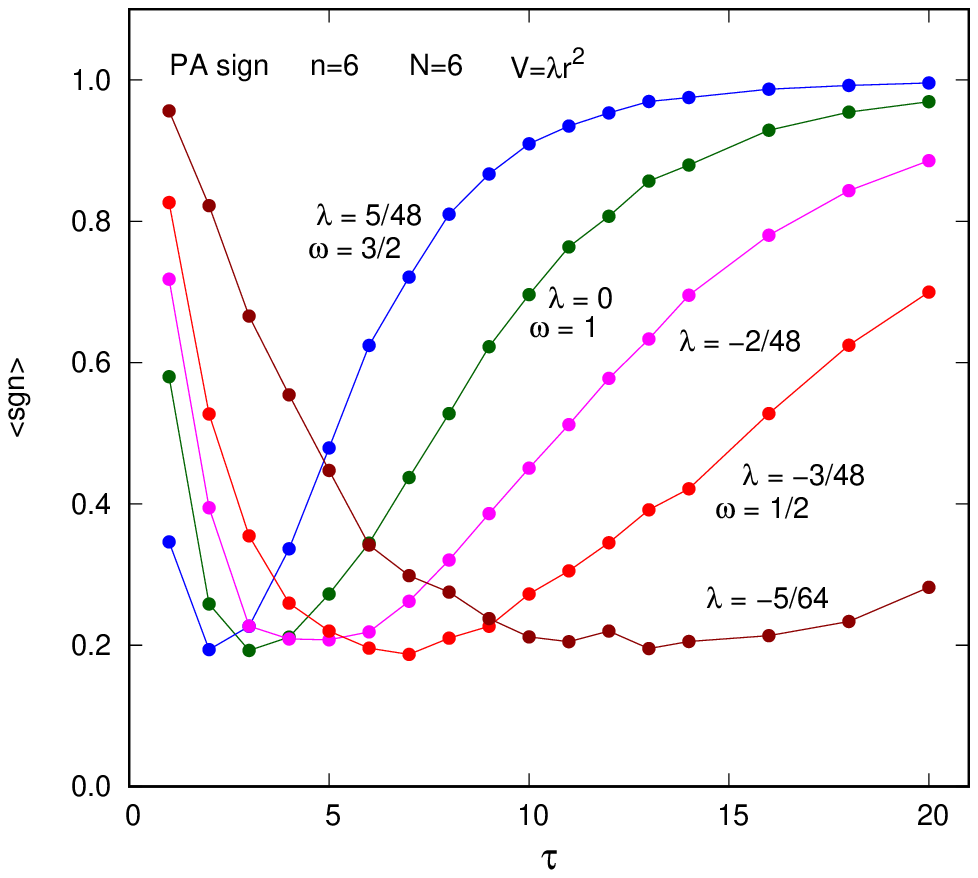}
	\includegraphics[width=0.49\linewidth]{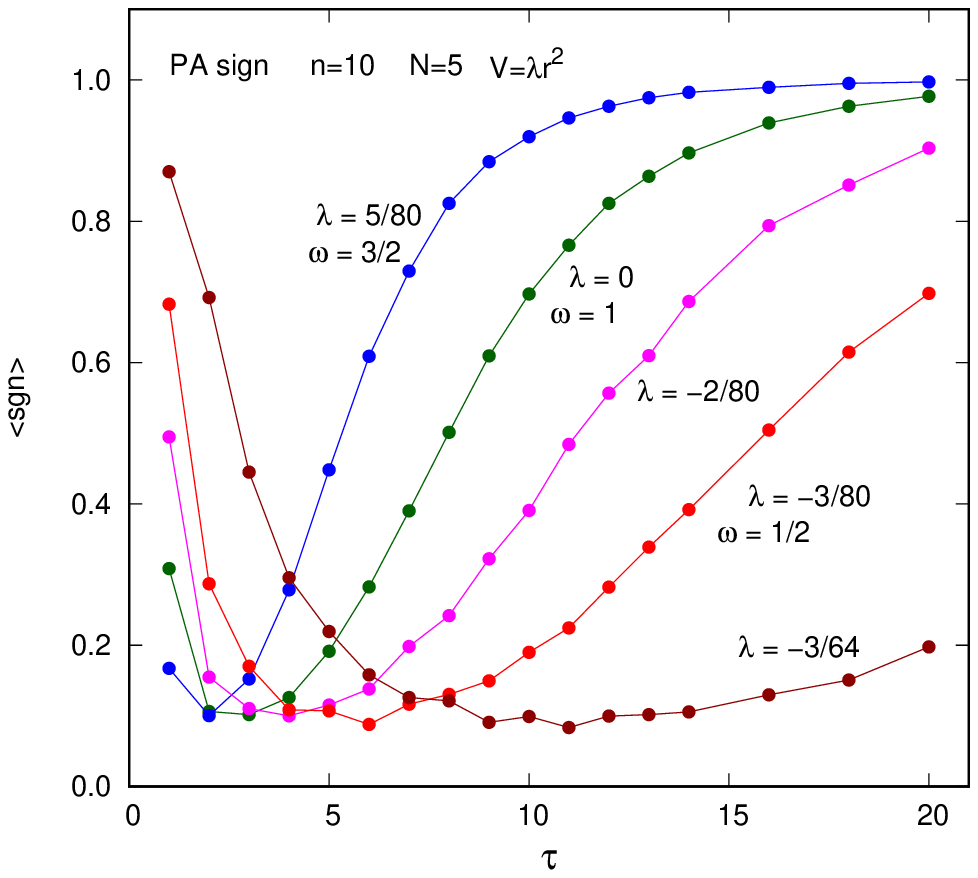}
	\caption{ (color online) 
		The change in the average sign as a function of the strength of the pairwise harmonic interactions for six and ten fermions in a 2D harmonic oscillator. 	}
	\la{n610sign}
\end{figure}

For our main interest in large quantum dots with repulsive pairwise forces, we set $\lm=-3/(8n)$ so that $\w=1/2$
for all $n$ values. The PA Hamiltonian energy for $n=6$ and $n=10$ are shown in Fig.\ref{neg610pa}.
In contrast to the attractive case, for $n=6$, H8 is affected by the sign problem with large error bars at $\tau<10$ but has no problem of reproducing the analytical result at $\tau>10$. For $n=10$, H6 and H8 have sign problems below $\tau<10$ and $\tau<15$ respectively but none above those values.

\begin{figure}[t]
	\includegraphics[width=0.49\linewidth]{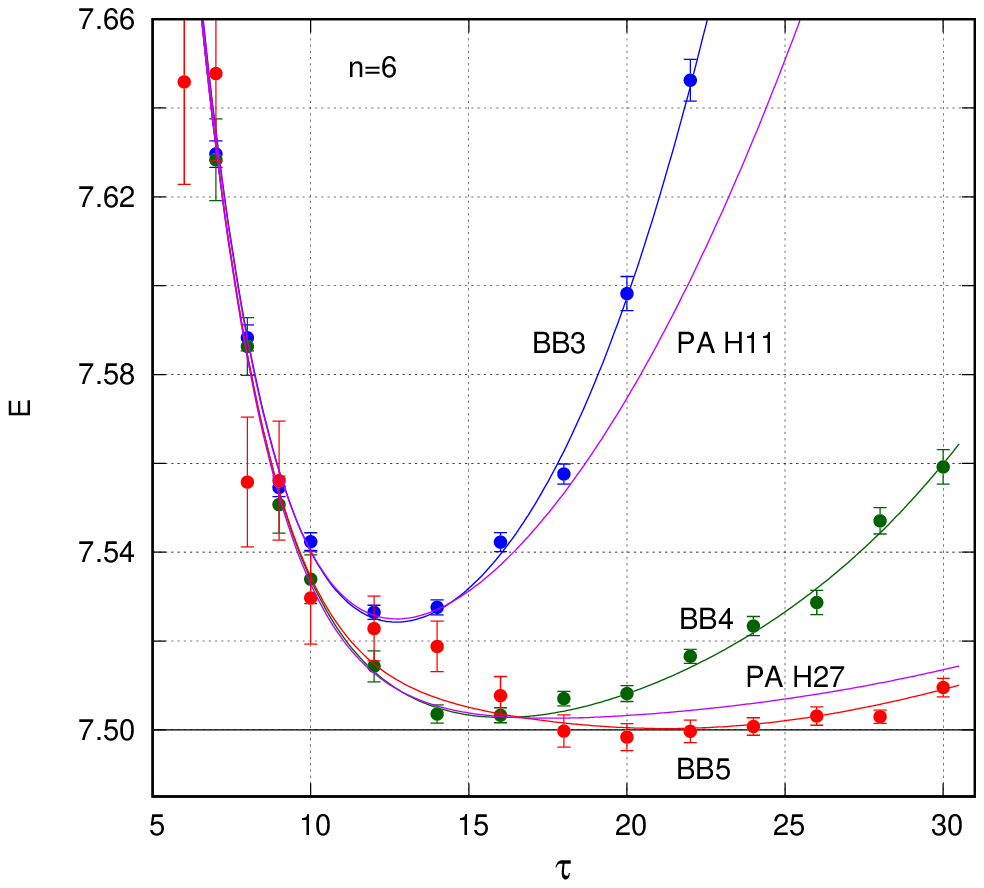}
	\includegraphics[width=0.49\linewidth]{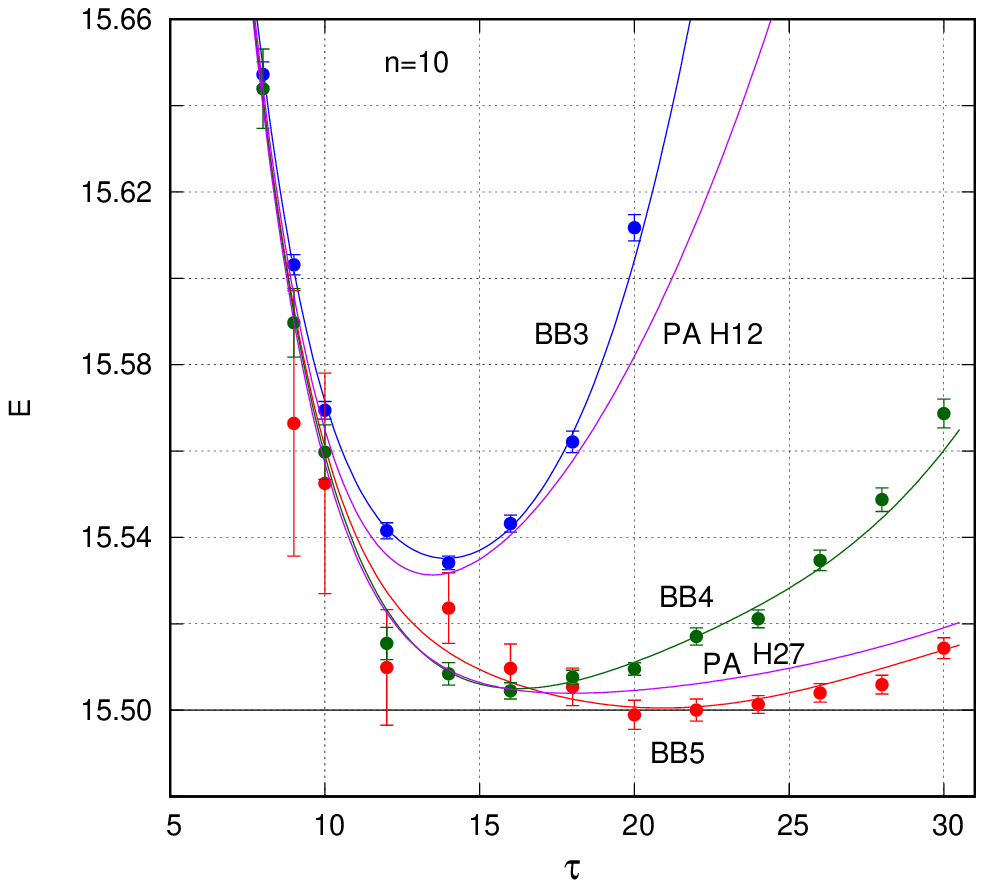}
	\caption{ (color online) 
		Hamiltonian energies for the same $n=6$ and 10-fermion system as in Fig.\ref{neg610pa}, 
		but solved by three fourth-order propagators BB3, BB4, BB5 as described in Appendix \ref{falg}. The smooth curves are obtained by evaluating 
		(\ref{hhw}) numerically without knowing its analytical form.
	}
	\la{neg610}
\end{figure}

These observations can now be understood by examining PA's average sign $s=\langle sgn \rangle$ 
as a function of interaction strength in Fig.\ref{n610sign}. 
One immediately notes that: 1) Attraction ($\lm>0$) reduces the overall sign problem as compared to the non-interacting $\lm=0$ case, except where $\tau<2$. This explains the absence of sign problems in Fig.\ref{enghar}.
2) Repulsion ($\lm<0$) reduces the sign problem at small
$\tau$ but broadens the sign minimum at large $\tau$. 3) The sign minimum with interaction is never lower than that of the non-interacting case, {\it i.e.}, pairwise interactions do not worsen the sign problem, 
only shift the sign minimum to a different $\tau$ location. 

This suggests that for strong
repulsion, one should start at small $\tau$ and equilibrate toward the ground state at large $\tau$,
as done in Ref.\onlinecite{chin15}. For weak to moderate repulsion, it would best to search for the energy minimum at large $\tau$, where the sign problem is less severe for closed-shell states. 
However, for PA this strategy can only work at large $N$, such as H6 and H8 
for a small number of fermions in Fig.\ref{neg610pa}. 
For large $n$, one must keep $N$ small to control the sign problem.

With repulsive interactions, BB3 can no longer reproduce the exact ground state energy as in the non-interacting case. However, the energy can be greatly improved with BB4 and nearly exactly reproduced by BB5. For up to five beads,
the sign problem for $n=10$ is still manageable with a sign minimum of $\approx 0.1$, similar to the PA case in the right panel of Fig.\ref{n610sign}. Moreover,
as shown in Fig.\ref{neg610}, BB3 is comparable to PA H11 for $n=6$ and comparable to PA H12 for $n=10$. BB4 is comparable to PA H27 for both cases. These two algorithms are therefore nearly four and seven times as efficient as PA. 
BB5 is nearly exact, with energy minima of 7.498(3) for $n=6$ at $\tau=20$  and 15.499(3) for $n=10$ also at $\tau=20$.

\begin{figure}[t]
	\includegraphics[width=0.49\linewidth]{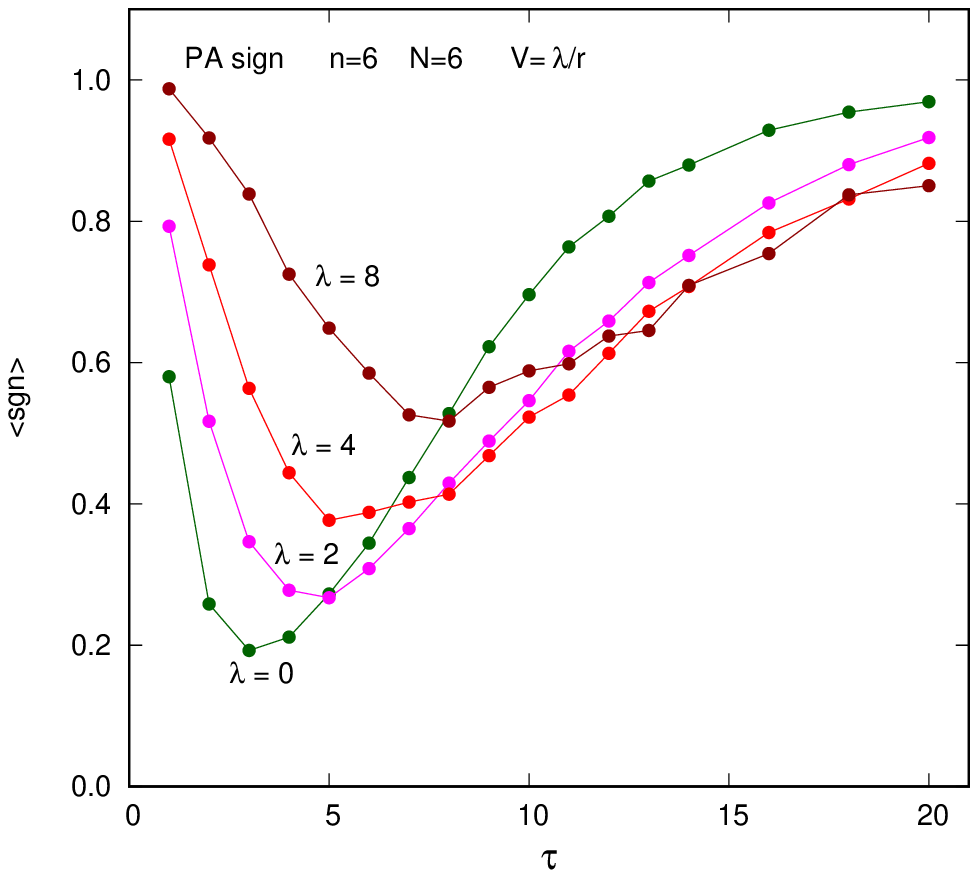}
	\includegraphics[width=0.49\linewidth]{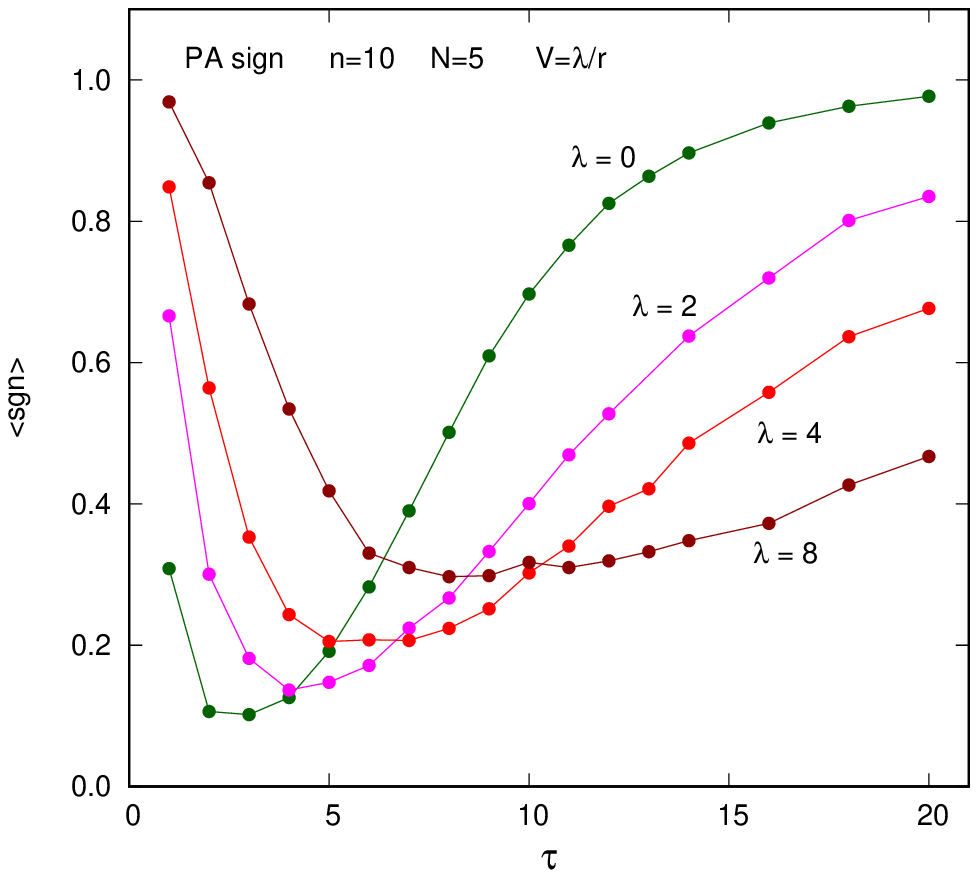}
	\caption{ (color online) 
		The change in the average sign as a function of the strength of the pairwise Coulomb repulsion for six and ten fermions in a 2D harmonic oscillator. 	}
	\la{n610rm1sign}
\end{figure} 

For the Coulomb repulsion, the average sign behaves even better. As shown in Fig.\ref{n610rm1sign} any Coulomb repulsion  lifts the sign minimum above that of the non-interacting case. However, as in the harmonic interaction case, strong repulsion greatly reduces the sign problem at small $\tau$, but suppresses the sign average at large $\tau$, thereby nullifying the closed-shell advantage.

\section{Spin-balanced quantum dots}
\la{sbqd}

\begin{figure}[t]
	\includegraphics[width=0.49\linewidth]{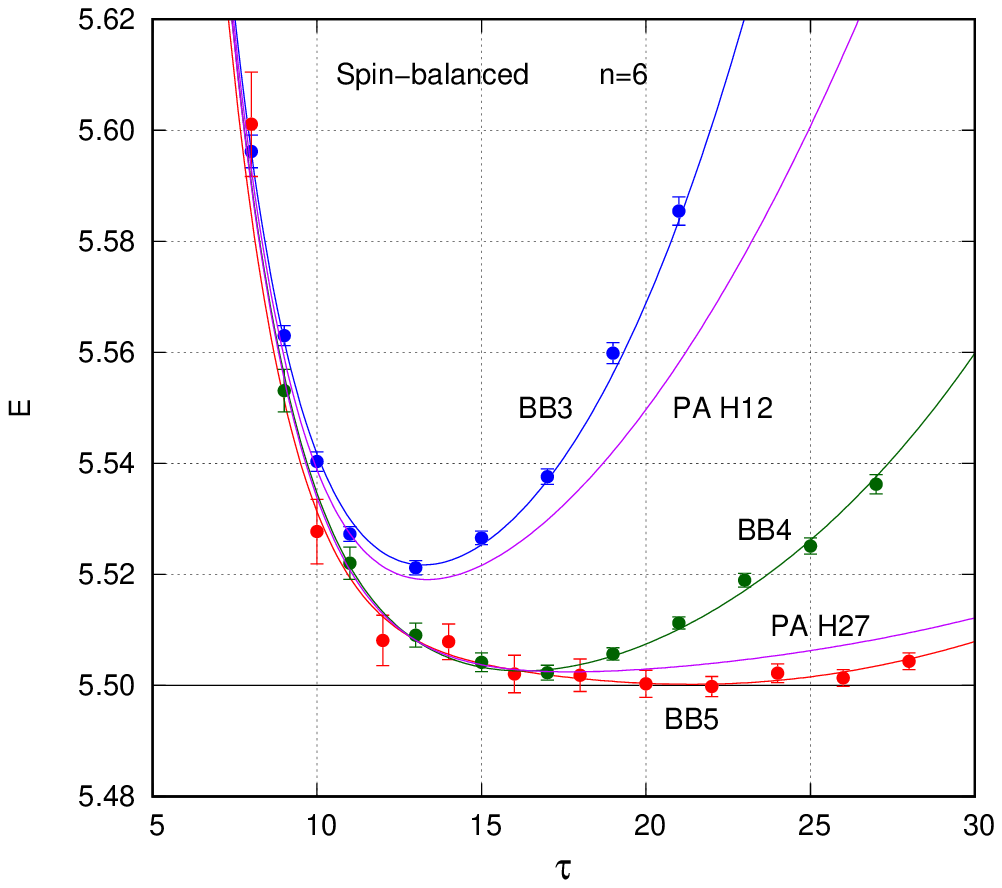}
	\includegraphics[width=0.49\linewidth]{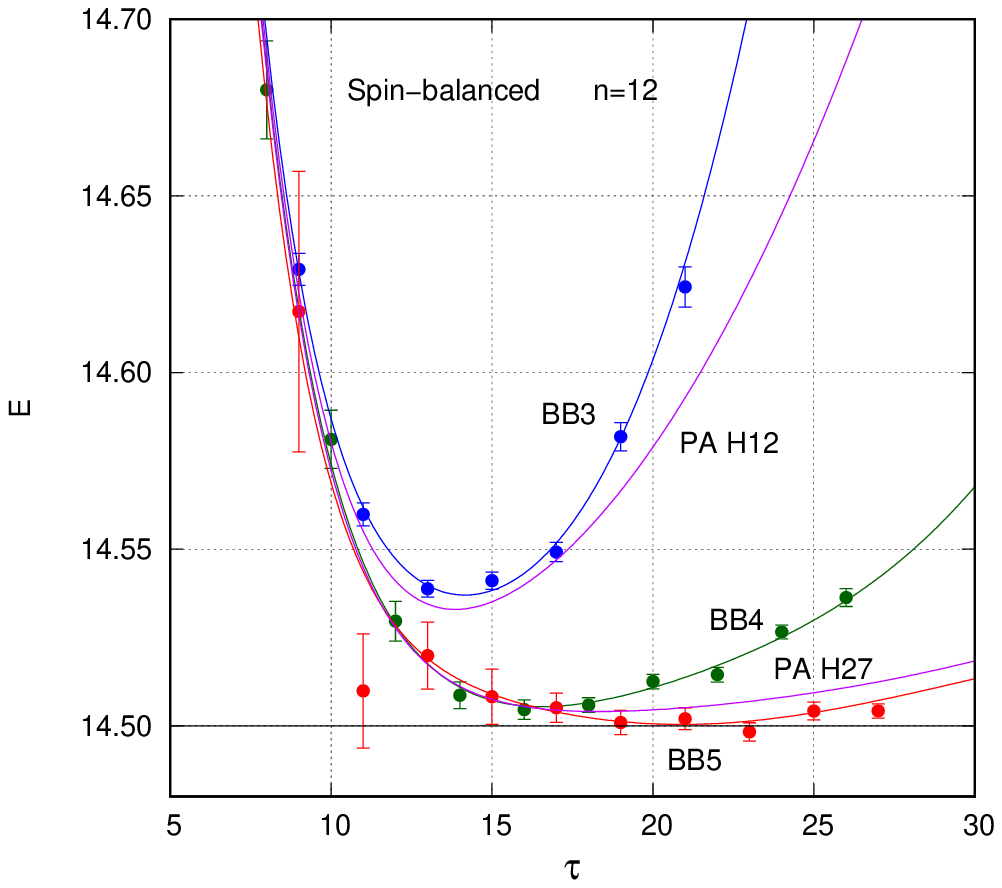}
	
	\includegraphics[width=0.49\linewidth]{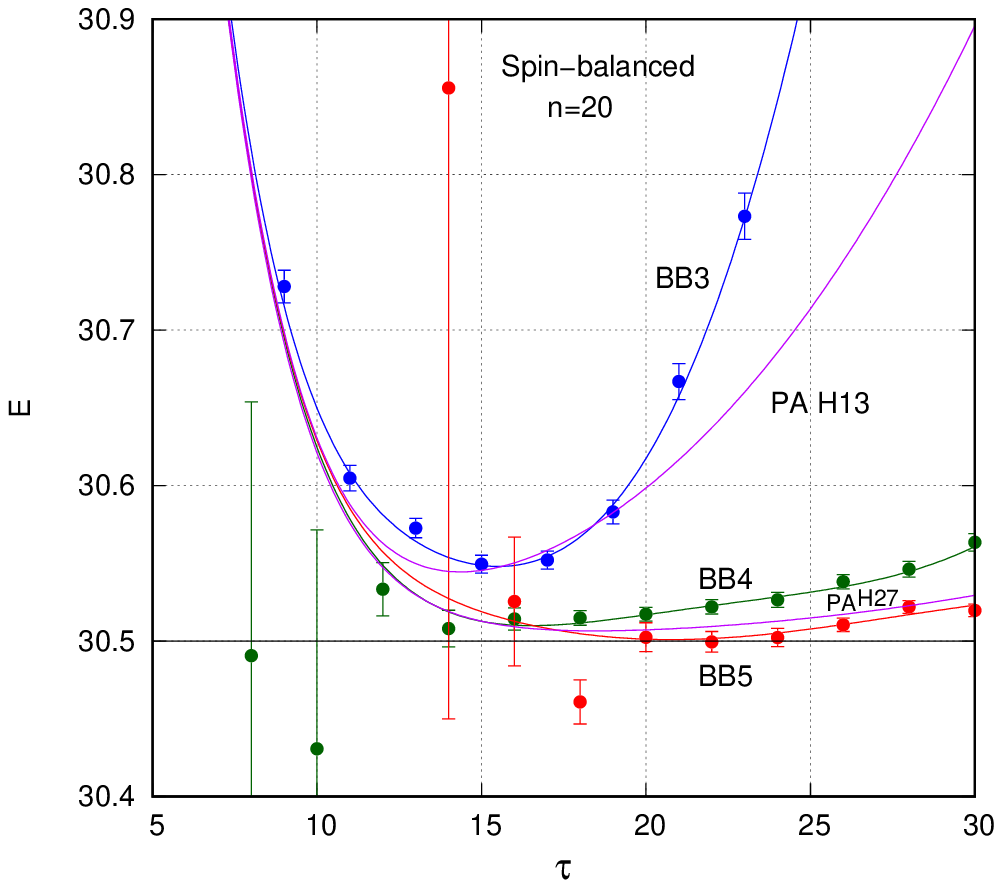}
	\includegraphics[width=0.49\linewidth]{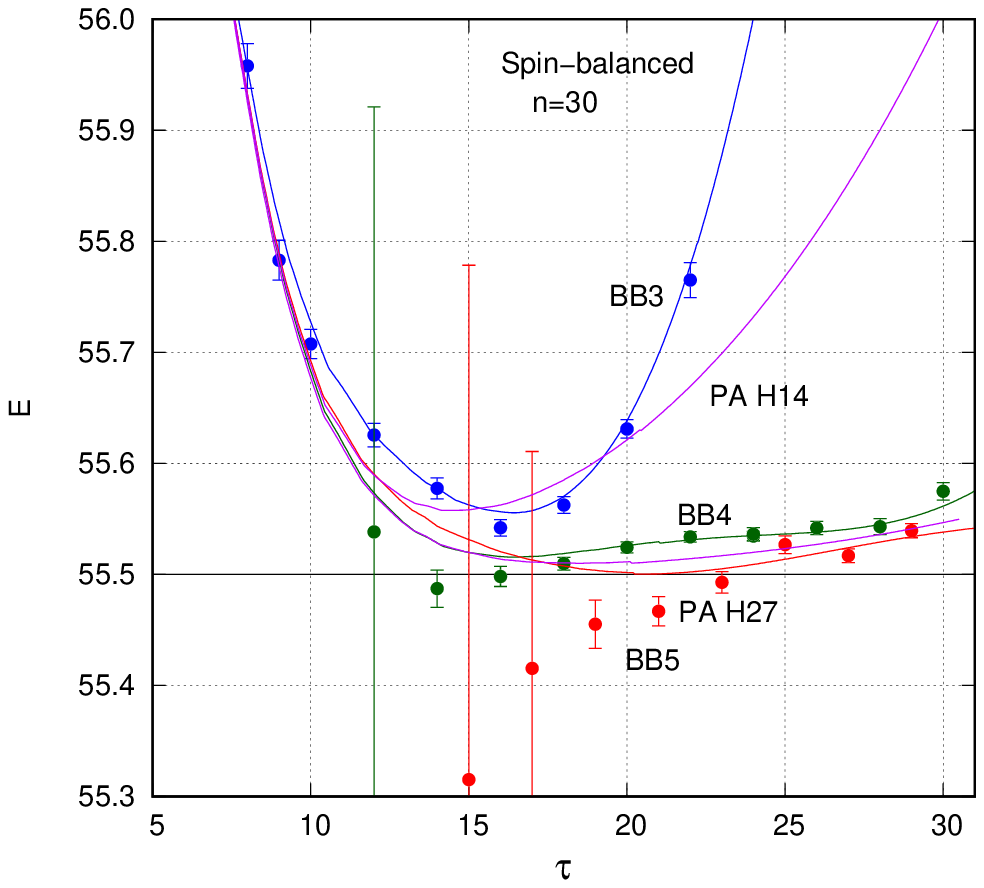}			
	\caption{ (color online) 
	   Spin-balanced $n=6,12,20,30$ Hamiltonian energies with pairwise repulsive harmonic
	   potential $\lm(\br_i-\br_j)^2$ and with $\lm$ adjusted so that $\w=1/2$ for all $n$. 
	The data points are PIMC calculations and smooth curves are analytical results given by (\ref{hww}).
	See text for details.
			}
	\la{negopt}
\end{figure}

In previous sections, we have studied spin-polarized (or spinless) fermions using 
a single determinant free propagator.
Consider now spin-balanced, unpolarized quantum dots with  $n_{\uparrow}=n_{\downarrow}=n/2$, 
where the anti-symmetric free propagator used,
\ba
G_0(\x^\prime,\x,\dt)&=&
\langle \x^\prime|\e_\ca^{ -\dt\hT}|\x\rangle\nn\\
&=&\frac1{n_{\uparrow}!}{\rm det}\left(\frac1{(2\pi\dt)^{D/2}}
\exp\left[ -\frac1{2\dt}( \br_{\uparrow i}^\prime-\br_{\uparrow j} )^2\right] \right)\nn\\
&&\quad\times\frac1{n_{\downarrow}!}{\rm det}\left(\frac1{(2\pi\dt)^{D/2}}
\exp\left[-\frac1{2\dt}(\br_{\downarrow i}^\prime-\br_{\downarrow j})^2\right] \right),
\ea
is a product of two determinants. The
analytical (or numerical) Hamiltonian energy can be computed by modifying (\ref{hhw}) to
\ba
E_n^H(\tau)
&=&  \frac12(\ga+\frac1{\ga}) E_1(Nu)+\frac{\w}2(\frac{\ga^*}{\w}+\frac{\w}{\ga^*})\Bigl(2E_{n/2}(Nu^*)-E_1(Nu^*)\Bigr).
\la{hww}
\ea
Since we have the analytical energies $E_3(w)$, $E_6(w)$ and $E_{10}(w)$, we can compute from above, the numerical
energies for spin-balanced systems of $n=6,12$, and 20. For $n=30$, one can again compute
$E_{15}(\tau)$ stochastically by using the exact but non-interacting propagator.
For repulsive forces such that $\w=1/2$ is the same for all
$n$, we compare PIMC data in Fig.\ref{negopt} with their analytical predictions.
The spin-balanced case is generally more difficult than the spin-polarized case. This is because even if the two spin-polarized subsystems can be solved exactly, one still has to account for pairwise interactions between the two subsystems. For harmonic interactions, as seen in Fig.\ref{negopt}, the optimized fourth-order algorithms are equally effective in solving spin-balanced cases. However, for $n=20$, the sign problem for BB5 is very
severe below $\tau<20$, where some data points with very large error bars are outside of the plotting range. For $n=30$, the sign problem is too severe for BB5, but BB4 can still give excellent estimate of the ground state energy.

For the Coulomb interaction, spin-polarized quantum dots have been studied extensively in Ref.\onlinecite{chin15}. 
For spin-balanced quantum dots, most published works use an alternative parametrization of the
 Schr\"odinger equation 
 \be
 (-\frac1{2}\sum_{i=1}^n\nabla_i^2
 +\frac12 \sum_{i=1}^n\w^2{\br_i}^2+\sum_{j>i}\frac1{|\br_i-\br_j|})\psi=E_{\w}\psi,
 \la{sch2}
 \ee
which is related to the interaction form  ${\lm}/{|\br_i-\br_j|}$ used here via
\be
\lm=\sqrt{ \frac1{\w} }\qquad {\rm and}\qquad E_\w=E/\lm^2.
\ee 
For the ease of comparing with published works, 
we will use $\w$ instead of $\lm$ as the coupling strength and just denote $E_\w$ as $E$ below. (The $\w$ here has nothing to do with the same symbol used in Eq.(\ref{omg}). This should not cause confusion because
we will only consider the Coulomb interaction from this point forward.) 

\begin{figure}[t]
	\includegraphics[width=0.49\linewidth]{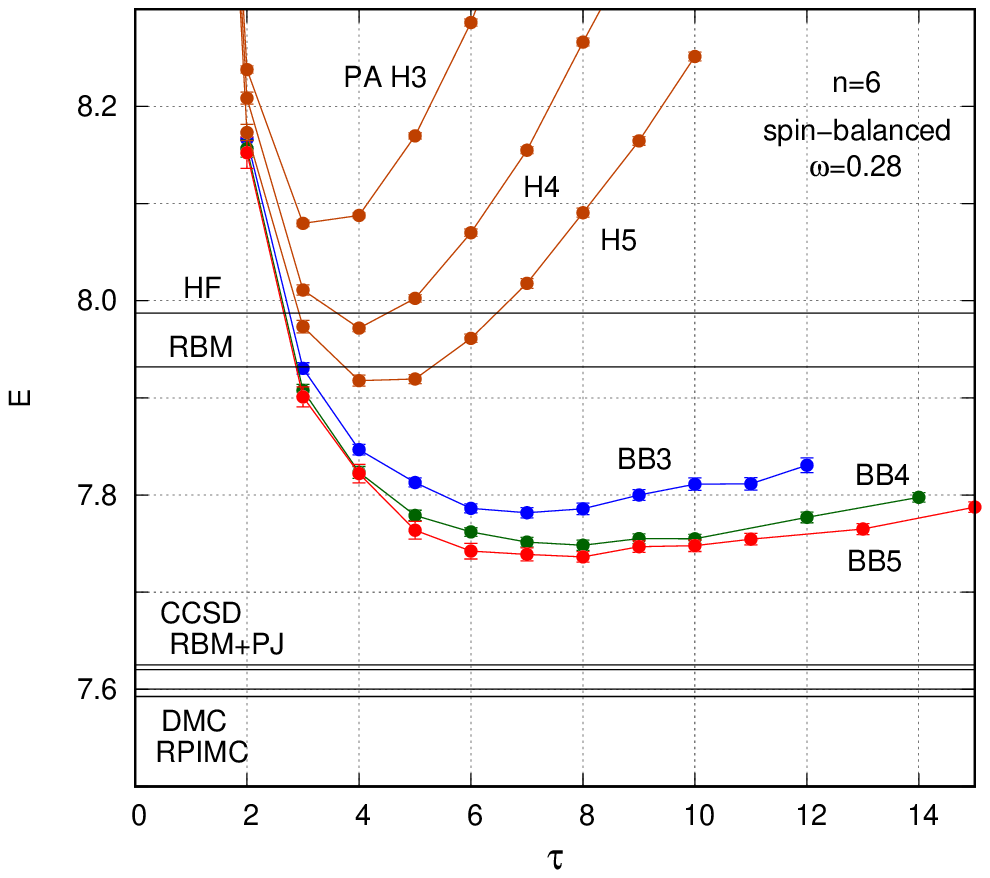}
	\includegraphics[width=0.49\linewidth]{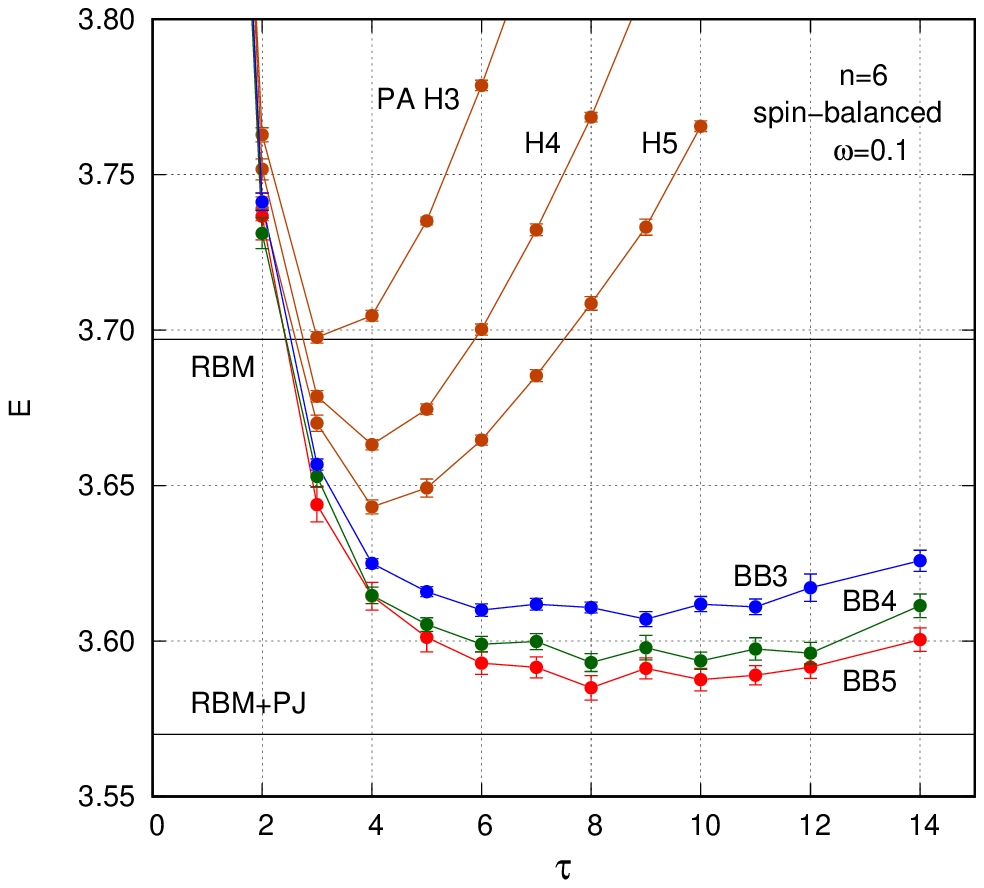}
	
	\includegraphics[width=0.49\linewidth]{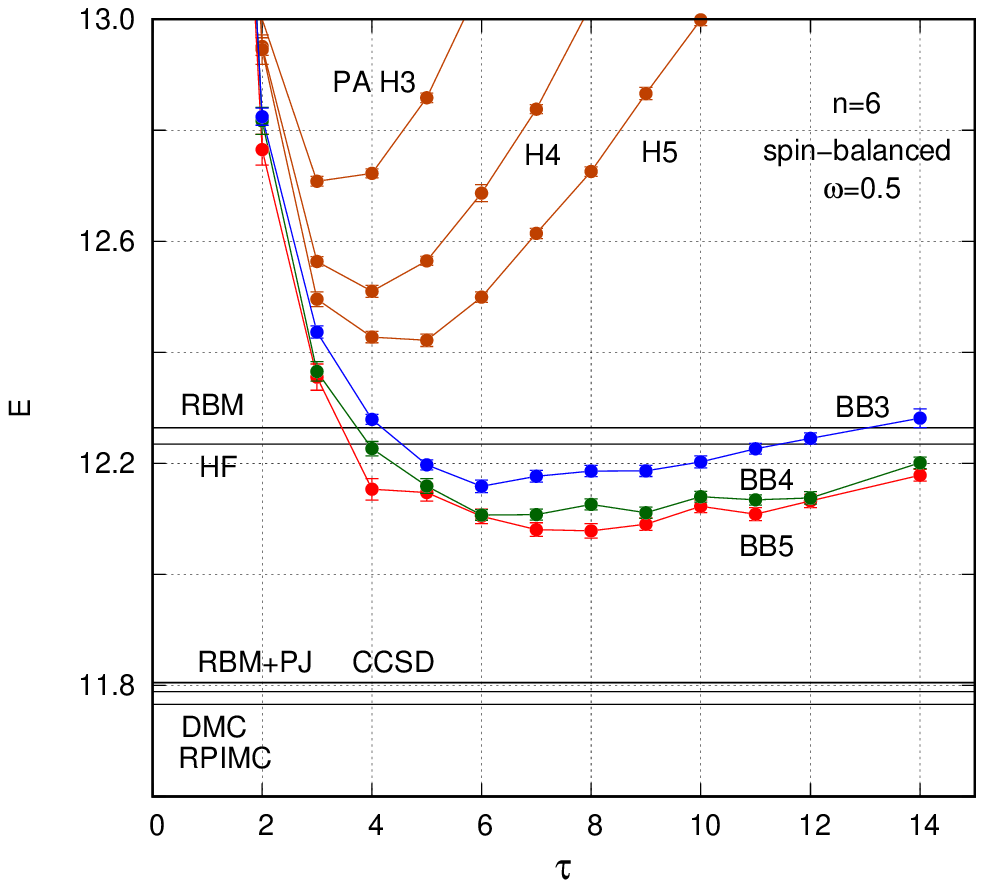}
	\includegraphics[width=0.49\linewidth]{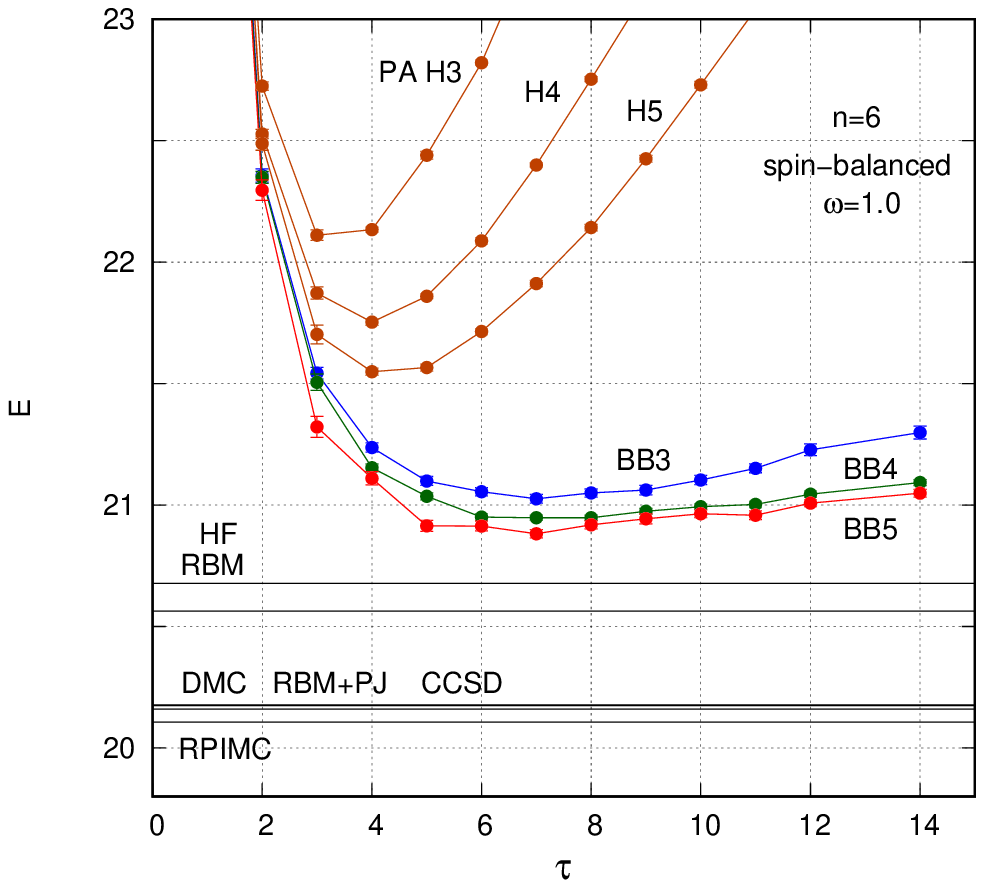}
	\caption{ (color online) 
		The energies of $n=6$ spin-balanced quantum dots at various values of $\w$ as compared with many published
		results. See text for details. 	}
	\la{n6}
\end{figure}

The energy results for $n=6$ are shown in Fig.\ref{n6}. The top-left panel is for $\w=0.28$, the coupling closest to the
experimental situation. Indicated by horizontal lines are the ground state energies obtained by the method of
Hartree-Fock (HF), Couple-Cluster with Single and Double excitation (CCSD), Diffusion Monte Carlo (DMC) 
by Lohne {\it et al.}\cite{loh11}, the Restricted-path Path Integral Monte Carlo (RPIMC) by Kyl{\"a}np{\"a}{\"a} and Esa R{\"a}s{\"a}nen\cite{kyl17}, and the Restricted Boltzmann Machine (RBM) and RBM plus Pad\`e-Jastrow (RBM+PJ) neural 
network results by Nordhagen {\it et al.}\cite{nor23}.

The first surprise is that since both RBM and HF are uncorrelated wave functions, it is not clear how RBM
can have energy lower than the optimized single-particle wave functions of HF. Second, the HF energy
can be bested by PA's H4 and both by H5. The BB3, BB4, BB5 results are, as expected, dramatically lower than 
those of PA. However, BB5's minimum 7.736(5) at $\tau=8$, remains 1.8\% above the {\it extrapolated} ground state energy 7.5927(2) of RPIMC and 1.5\% above 7.6203(2) of RBM+PJ.

The situation gets worse as one weakens the coupling to $\w=0.5$ in the lower-left panel.
Here, HF is lower than RBM and H5. BB5's energy of 12.08(1) at $\tau=8$ is 2.7\% above RPIMC's 11.7659(4)
and 2.3\% above RBM-PJ's 11.80494(7). Weakens the coupling to $\w=1.0$ in the lower-right panel shows that
HF is below BB5 and BB5's minimum energy of 20.88(2) at $\tau=7$ is 3.5\% above RBM-PJ's 20.1773(1).
On the other hand, if one strengthens the coupling to $\w=0.1$ (only done in Ref.\onlinecite{nor23}),
as shown in the upper-right panel, then even H3 is comparable to RBM and BB5's minimum energy of 3.585(4) 
at $\tau=8$ is only 0.42\% above RBM+PJ's 3.5700(2).

This systematic effect is not due to the sign problem, which is manageable for $n=6$ and $N$ at most 5.
This is related to the challenge that faces PIMC in dealing with the singular Coulomb potential, 
where the exact wave function has cusps. 
In DMC and RBM+PJ, the cusp condition is built into the trial function. For PIMC, the cusp condition can
only be approximated at large $N$, but this then runs into the sign problem. The present work, which seeks to 
circumvent the sign problem by restricting $N<5$, the cusp condition can only be avoided at large couplings, 
where the strong Coulomb repulsion would keep the particles far apart, away from the cusp of the wave function.
This explains the success of Ref.\onlinecite{chin15}, which unwittingly, but fortunately, used $\lm=8$. 

\begin{figure}[htb]
	\includegraphics[width=0.49\linewidth]{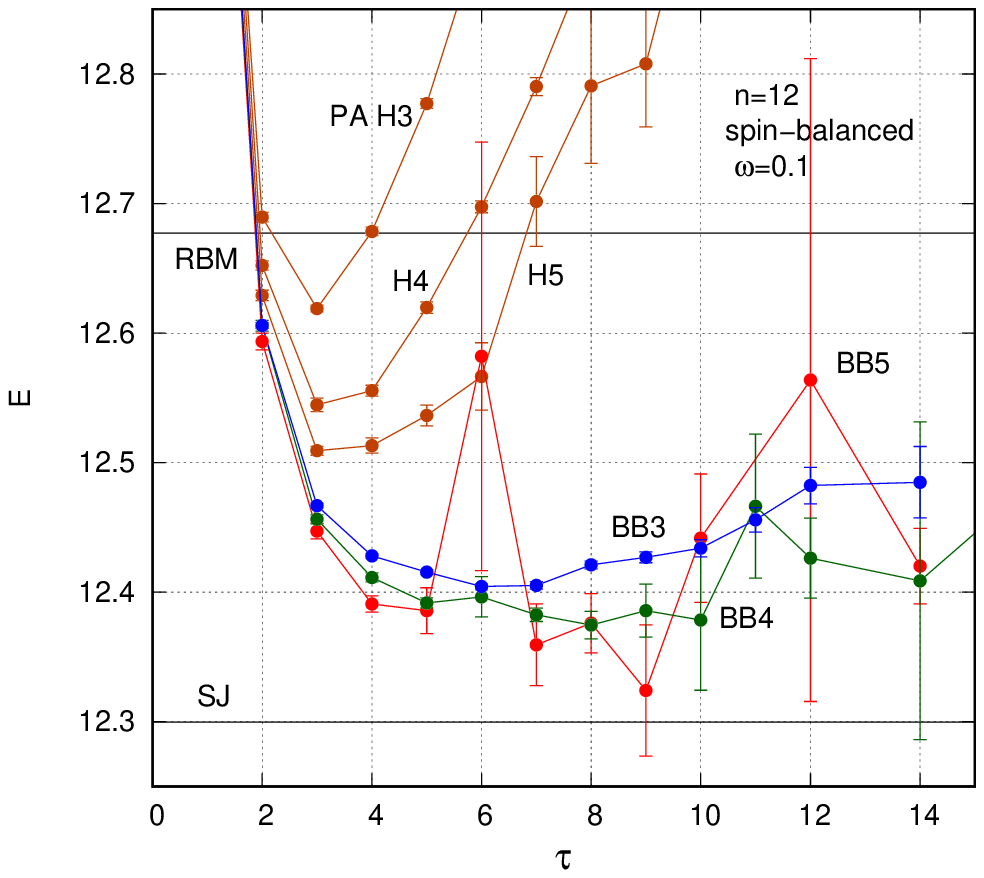}
	\includegraphics[width=0.49\linewidth]{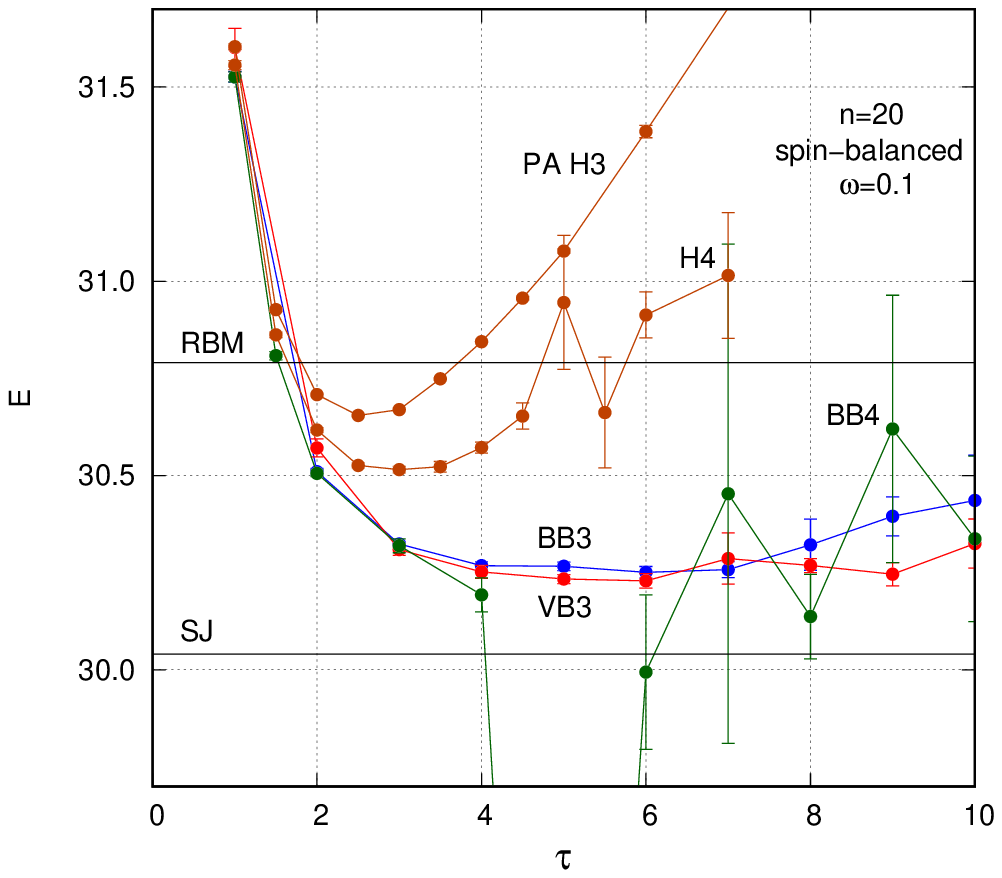}
	\caption{ (color online) 
		Energies of larger $n=12$ and $n=20$ unpolarized quantum dots, where the sign problem begins to 
		force out fourth-order propagators. Algorithm VB3 is described in Sect.\ref{lqd}.}
	\la{n12}
\end{figure}

For $n=12$ and 20 at $\w=0.1$, similar results are shown in Fig.\ref{n12}. For these larger quantum dots, 
Ref.\onlinecite{nor23}'s Slater-Jastrow (SJ) based neural network has lower energies. One sees that for $n=12$
BB5 is barely calculable with large sign errors and cannot be used for $n=20$. Nevertheless its minimum energy of $12.32(5)$ for $n=12$ at $\tau=9$ is only 0.1\% above Ref.\onlinecite{nor23}'s SJ energy of 12.29962(9). For $n=20$, BB4 is
barely doable and its energy of 30.19(4) at $\tau=4$ is 0.5\%  above  Ref.\onlinecite{nor23}'s SJ energy of 30.0403(2).
For $n=30$ and larger, not even BB3 is stable. This is consistent with previous energy results on 
spin-polarized\cite{chin15} quantum dots.

\section{large quantum dots}
\la{lqd}

From Fig.\ref{n12} one sees that the sign problems for BB5 and BB4 are more severe 
than their PA counter parts H5 and H4. Evidently, the gradient potential ($\approx 1/r^4_{ij}$) 
for the pairwise Coulomb interaction, especially at large $n$, 
has exacerbated the sign problem. At small $N$, PA's sign problem seemed more manageable, but its energy is not
low enough.

\begin{figure}[thb]
	\includegraphics[width=0.49\linewidth]{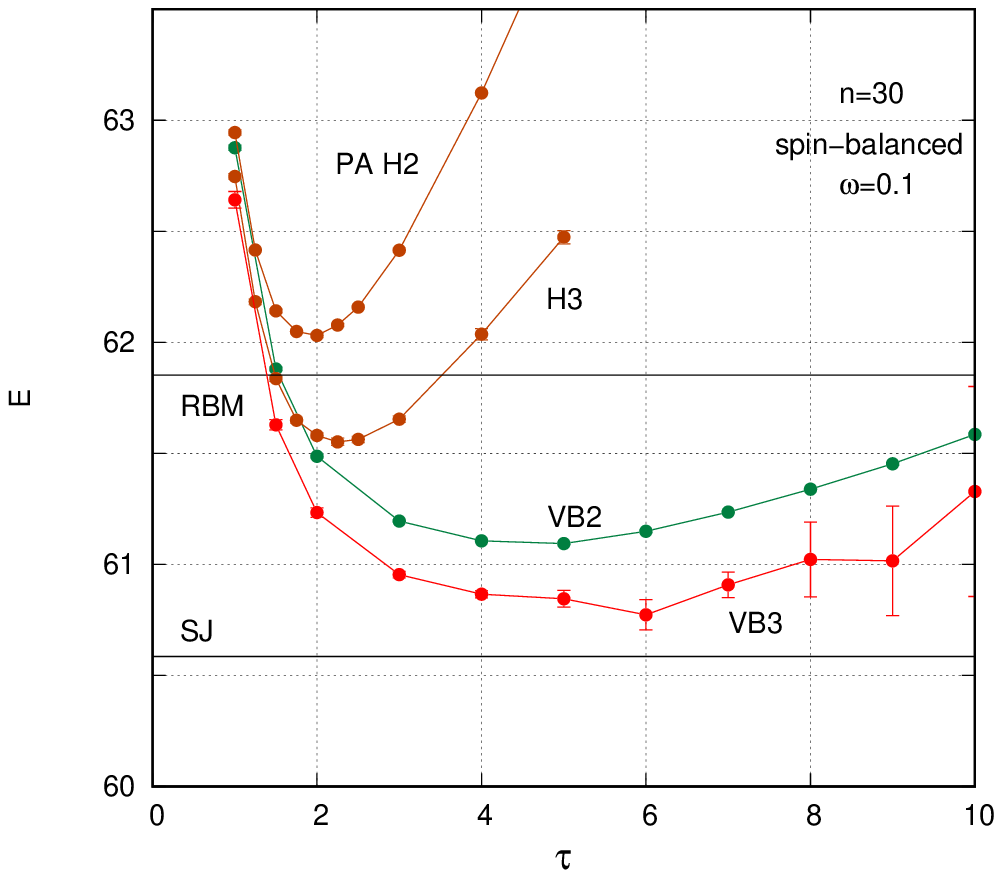}
	\includegraphics[width=0.49\linewidth]{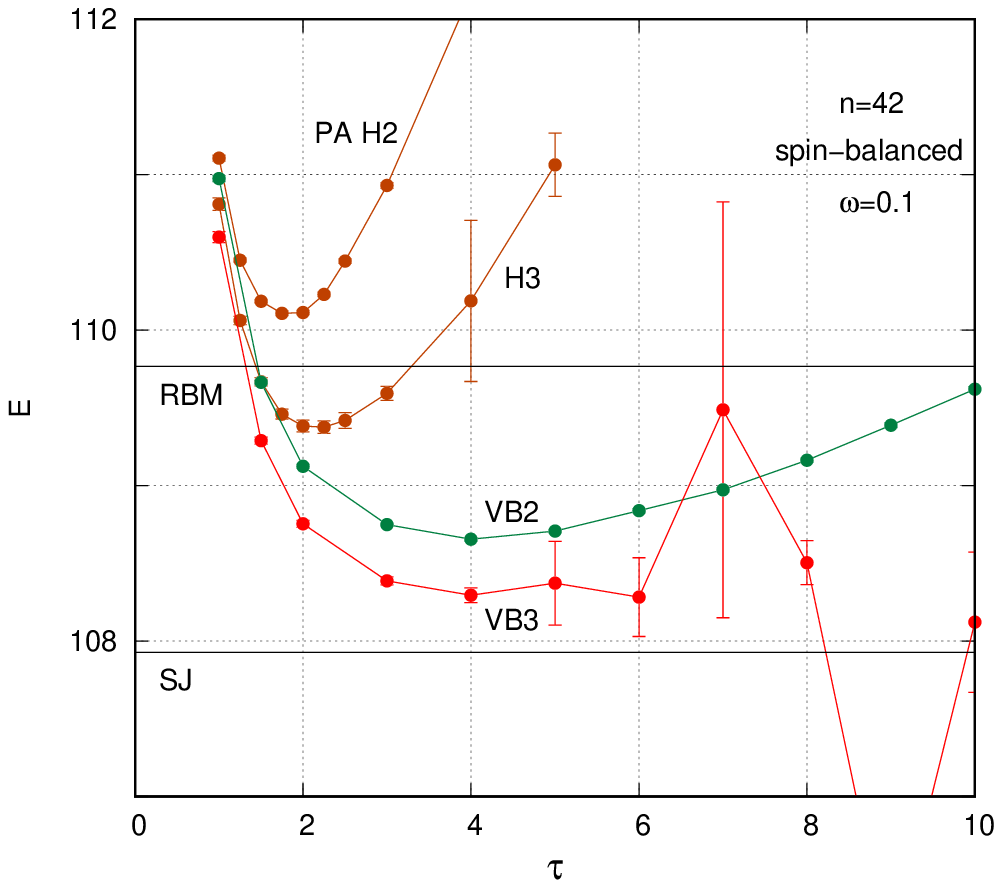}
	
	\includegraphics[width=0.49\linewidth]{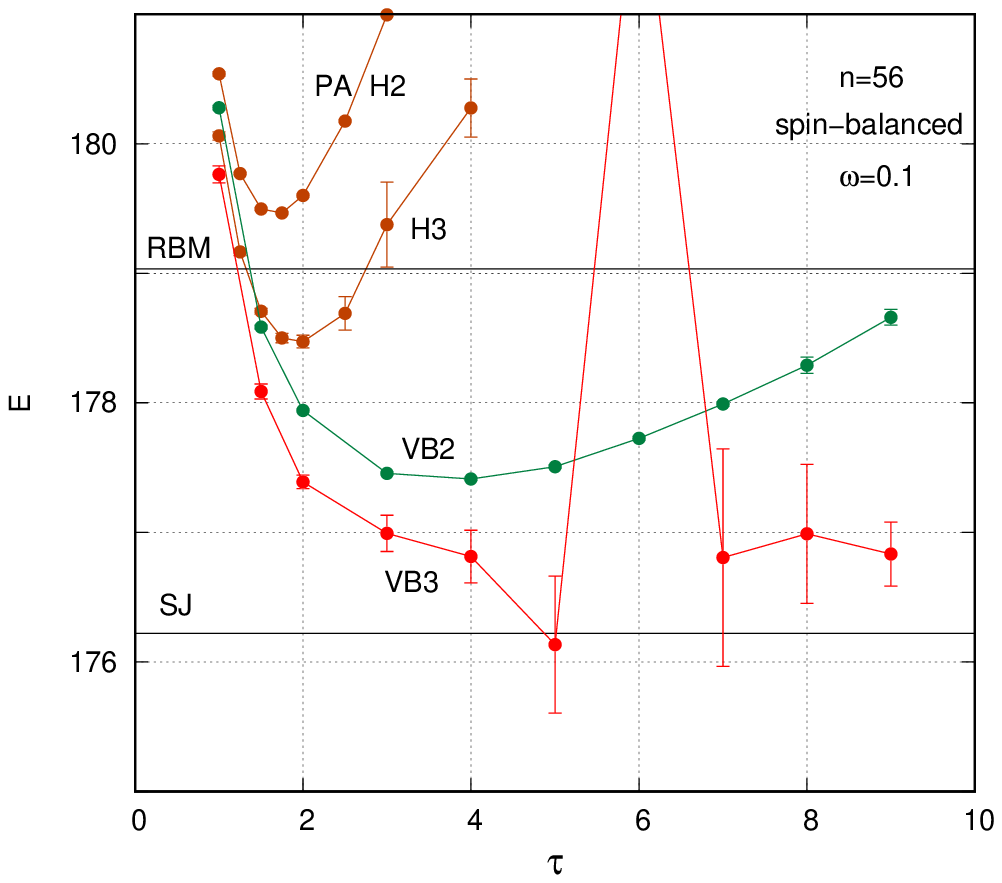}
	\includegraphics[width=0.49\linewidth]{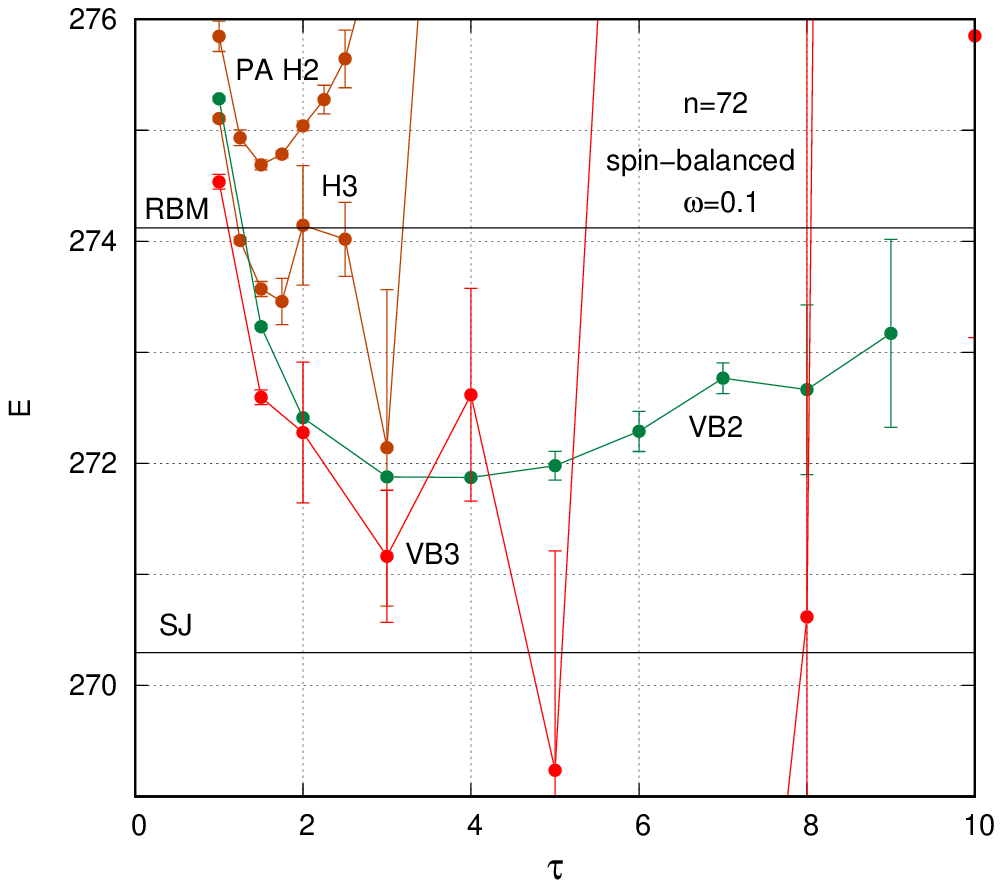}
	\caption{ (color online) 
		Large quantum dot calculations using Variable-Bead propagators VB2 and VB3. See text for details.	}
	\la{n30}
\end{figure}

      The success of fourth-order propagators is also due to the fact that they have parameters that
can be optimized to lower the energy. Surprisingly, such parameters also exist for the PA algorithm.
The PA 2-bead propagator given by
\be
\hG_2(\dt)=\e^{ -\frac12\dt\hV}\e^{ -\dt\hT}\e^{ -\dt\hV}\e^{ -\dt\hT}\e^{ -\frac12 \dt\hV},
\la{pag2}
\ee
has a hidden parameter $v_1$, 
\be
\hG_2(\dt)=\e^{ -\frac12 v_1\dt\hV}\e^{ -\dt\hT}\e^{ -(2-v_1)\dt\hV}\e^{ -\dt\hT}\e^{ -\frac12 v_1 \dt\hV}.
\la{pag2h}
\ee
As long as the propagator remains left-right symmetric, it is second-order.
The case of $v_1=1$ reproduces the original PA of (\ref{pag2}) and $v_1=2$ gives PA at step size $2\dt$.
Clearly then, PA can only be improved for $v_1<1$. In Fig.\ref{n30}, we show its energy for $n=30$, 42, 56 and 72 quantum dots at $v_1=0.25$. Its energy is dramatically lower than the unvarnished PA H2.
Its energy minima, recorded in Table \ref{tab1}, are midway between that of RBM and SJ for all $n$.
We shall refer to this as the Variable-Bead, VB2 propagator. 
VB2, like PA H2, has no sign problem.

\begin{figure}[t]
	\includegraphics[width=0.49\linewidth]{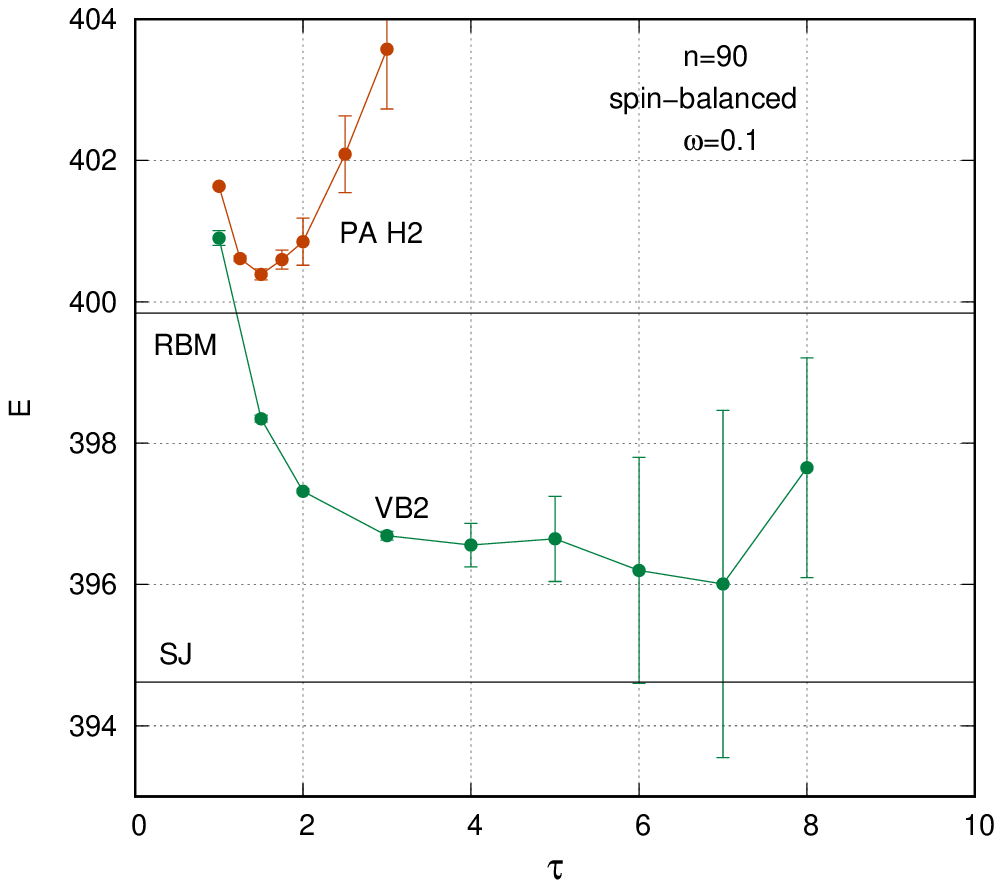}
	\includegraphics[width=0.49\linewidth]{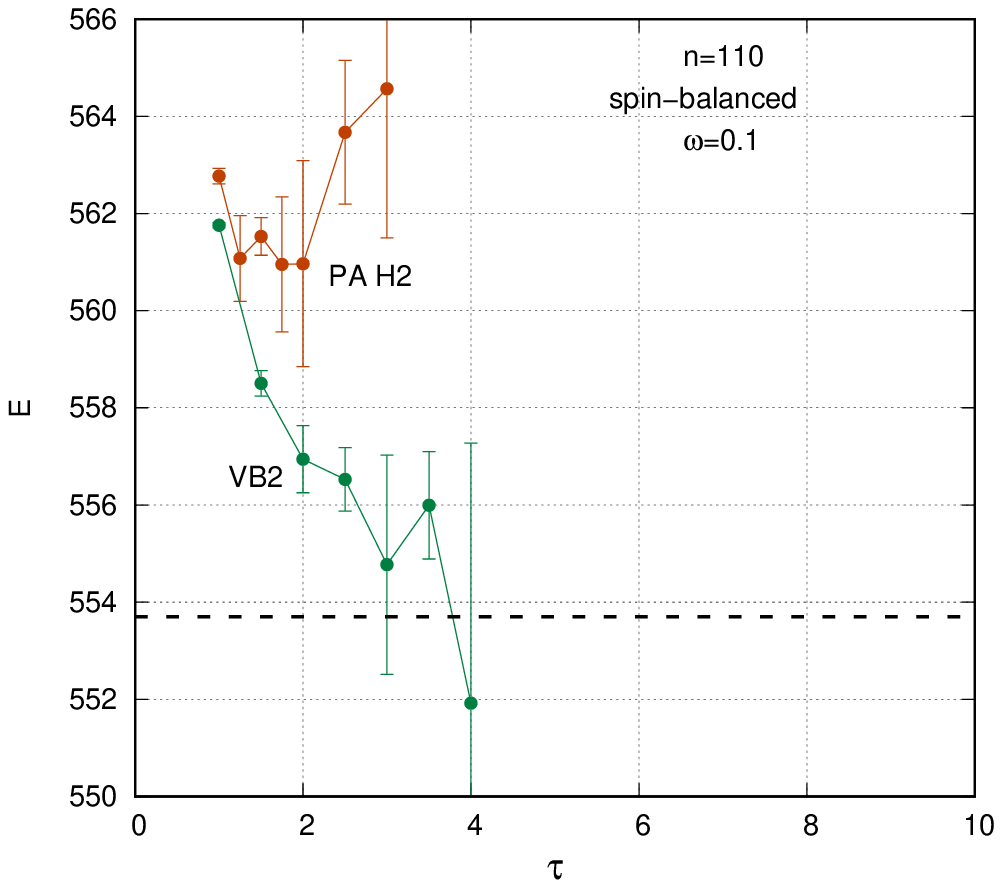}
	\caption{ (color online) {\bf Left:} 
		The Hamiltonian energy of a 90-electron unpolarized quantum dot
		as a function of imaginary time $\tau$. Algorithm VB2's energy of 396.69(6) at $\tau=3$, is only 0.5\%
		above Ref.\onlinecite{nor23}'s SJ neural network energy of 394.621(4). 
		{\bf Right:} The Hamiltonian energy for a 110-electron unpolarized quantum dot.
		Algorithm VB2's energy at $\tau=2.5$ is 556.5(6).}
	\la{n90}
\end{figure}

For 3-bead, one can choose the two free parameters $v_1$ and $t_1$ more concisely as follow:
\be
\hG_3(\dt)=\e^{ -\frac12(3-2 v_1)\dt\hV}\e^{ -t_1\dt\hT}\e^{ -v_1\dt\hV}\e^{ -(3-2t_1)\dt\hT}
\e^{ -v_1\dt\hV}\e^{ -t_1\dt\hT}\e^{ -\frac12(3-2 v_1)\dt\hV}.
\la{pag3h}
\ee
One recovers PA H3 at $v_1=1$, $t_1=1$, and $v_1=0$ collapses the propagator to that of PA at step size $3\dt$.
For $t_1=0.5$ and $v_1=1.4$, the minimum energy of this VB3 propagator for $n=30$, 42, 56, 72 at $\tau=6$, 4, 4, 3,
is shown in Fig.\ref{n30} and recorded in Table \ref{tab1}. All are only
only 0.3\% above Ref.\onlinecite{nor23}'s SJ energy. By contrast, VB2's energies at
$\tau=5$, 4, 4 and 4 are 0.8\%, 0.7\%, 0.7\% and 0.6\% above Ref.\onlinecite{nor23}'s SJ energy respectively.

Our error bars are much larger because each
calculation was done sampling at most $2\times 10^7$ particle-configurations on desktop computers.
From Fig.\ref{n12}, for $n=20$, VB3 is comparable to BB3, despite being only a second-order algorithm.
For $n=72$, VB3 has huge sign errors. For $n=90$, VB3 is no longer stable and only VB2 can be applied in
Fig.\ref{n90}. Nevertheless, VB2's energy is only 0.5\% higher than Ref.\onlinecite{nor23}'s SJ
energy. For $n=110$, also shown in Fig.\ref{n90}, there is no published energy for such a large quantum dot.
If VB2's energy remains about 0.5\% above the ground state energy, then the
ground state energy can be inferred to be approximately 553.7, plotted as a dash line in Fig.\ref{n90}.

The energy results in Fig.\ref{n90} took weeks to produce on multiple desktop computers. It is at the
limit of what personal computers can do.

\begin{table}
	\caption{Comparing this work's VB2 and VB3 spin-balanced quantum dot ground state energies 
		at coupling $\w=0.1$ with those of RBM and SJ neural networks\cite{nor23}.}
	\begin{center}
		\begin{tabular}{c r r r r }
			\colrule
			$n$ &\ \  {\rm RBM}\cite{nor23}& {\rm VB2 }\quad\quad & {\rm VB3}\quad\quad&{\rm SJ}\cite{nor23}\quad\quad \\
			\colrule
			30 &\qquad\ 61.853(2)   &\qquad\ \ 61.094(5)   &\qquad 60.77(7)   &\qquad 60.585(1)    \\
			42 &\qquad 109.767(7)   &\qquad 108.65(1)   &\qquad 108.29(5)  &\qquad 107.928(2)   \\
			56 &\qquad 179.035(8)   &\qquad 177.41(1)   &\qquad 176.8(2)\ \ &\qquad 176.221(1)  \\
			72 &\qquad 274.12(1)\ \ &\qquad 271.87(1)   &\qquad 271.2(6)\ \ &\qquad 270.296(3)  \\
			90 &\qquad 399.84(1)\ \ &\qquad 396.69(6)   &   $\cdots$\qquad\quad  &\qquad 394.621(4)  \\
			110 & $\cdots$\qquad\qquad  &  556.5(6)\ \  &   $\cdots$\qquad\quad  & $\cdots$ \qquad\qquad \\
			\colrule
		\end{tabular}
	\end{center}
	\label{tab1}
\end{table}

\section{Conclusions and future directions}
\la{con}

    In this work we have formulated a completely solvable PIMC model of harmonic fermions
with and without pairwise harmonic interactions where the Hamiltonian energy 
and the sign problem can be studied in details as functions of bead and fermion numbers.
The sign problem is primarily a property of the free fermion propagator, but
strong pairwise repulsion can reduce (or enhances) the sign problem at short (or long) imaginary time.
Most surprisingly, we found strong numerical evidence that there is no sign problem for 
closed-shell states in the large imaginary time limit. Moreover, we can show analytically, that
the way in which the first closed-shell state in dimension $D$ avoided the sign problem 
is exactly the $D$ dimensional generalization of the way in which the sign problem is avoided in one dimension. 
It would be of great interest, if one can also prove analytically, the absence of the sign problem in
higher closed-shell states.

This work also showed that fourth-order short-time propagators are
effective in determining ground state energies of less than 30 fermions with pairwise 
harmonic interactions or for sufficiently repulsive Coulomb forces. For spin-balance, unpolarized
quantum dots with 30 or more electrons, the pairwise gradient potential is too singular 
for fourth-order methods to be stable. However, this work found a new class of
Variable-Bead algorithms which yielded ground state energies that are only 0.5\% or less
above modern neural network results for up to 90 electrons. This was completely unexpected.
The efficiency of this class of algorithms should be further tested on other quantum systems.

     Path integral Monte Carlo is the granddaddy of modern neural network in that it is the only
traditional quantum Monte Carlo method with build-in ``hidden-layers". When PA is used, these layers
are completely known and fixed. The Variable-Bead propagators have hidden-layer parameters
that can be fine-tuned, making them similar to neural networks. However, these parameters
do not alter the fundamental Gaussian nature of PIMC, and therefore cannot overcome the fundamental
challenge of dealing singular potentials. Propagators such as (\ref{pag2h}) and (\ref{pag3h})
are actually ways of approximating the {\it square} of the wave function.
Perhaps one can learn from neural networks and replaces
the free propagator's Gaussian determinant by a more general type of determinant,
similar to the way Ref.\onlinecite{pfa20}'s neural network is parametrizing the exact wave function.
 
On the other hand, the VB2 and VB3 algorithms used here are much, much simpler than any fermion neural 
network\cite{nor23,pfa20}, suggesting that neural networks may also benefit from knowing 
the structure of PIMC in their search for the ground state wave function.       

\bigskip\bigskip\bigskip

\noindent  
{\bf AUTHOR DECLARATIONS}

\noindent 
{\bf Conflict of Interest}

The authors have no conflicts to disclose.

\noindent
{\bf Author Contributions}

Siu A. Chin: Conceptualization (lead); Formal analysis (lead);
Visualization (lead); Writing– original draft (lead).

\noindent 
{\bf DATA AVAILABILITY}

The data that supports the findings of this study are available
within the article and by inquiry to the author.

\appendix
\section {Needed integrals }
\la{aa}

The fundamental integral needed is
\be
z_n= \frac1{(2\pi\ka_N)^{nD/2}}
\int d\br_1d\br_2\cdots d\br_n \e^{-\mu_N(\br_1^2+\br_2^2+\cdots\br_n^2)}
\e^{-(\br_{12}^2+\br_{23}^2\cdots+\br_{n1}^2)/2\ka_N}.
\ee
This is the trace of the propagator
\be
G_n(\br_1,\br_{n+1})=G_1(\br_1,\br_2)G_1(\br_2,\br_3)\cdots G_1(\br_n,\br_{n+1})
\ee
where
\be
G_1(\br_i,\br_j)=\frac1{(2\pi\ka_N)^{D/2}}
\e^{-\mu_N\br_i^2/2}\e^{-\br_{ij}^2/2\ka_N}\e^{-\mu_N\br_j^2/2}.
\ee
This $G_1(\br_i,\br_j)$ can be regarded as a short-time propagator with
coefficients $\widetilde\mu_1$ and $\widetilde\ka_1$ given as
\be
\widetilde\mu_1=\mu_N\qquad{\rm and}\qquad \widetilde\ka_1=\ka_N.
\ee
Therefore, by contracting the fermion number $n$,
\be
G_n(\br_1,\br_{n+1})=\frac1{(2\pi\widetilde\ka_n)^{D/2}}
\e^{-{\widetilde\mu_n}\br_1^2/2}\e^{-(\br_1-\br_{n+1})^2/2{\widetilde\ka_n}}
\e^{-{\widetilde\mu_n}\br_{n+1}^2/2},
\ee
one has the trace
\ba
z_n&=&\frac1{(2\pi\widetilde\ka_n)^{D/2}}\int d\br_1 \e^{-{\widetilde\mu_n}\br_1^2}
= \frac1{(2\pi\widetilde\ka_n)^{D/2}}\left(\frac{\pi}{\widetilde\mu_n} \right)^{D/2}\nn\\
&=&\frac1{[2(\widetilde\za_n-1)]^{D/2}}=\frac1{[2\sinh(n\widetilde u/2)]^D}.
\ea
From the fundamental definition of $\widetilde u$,
\be
\cosh(\widetilde u)=\widetilde\za_1=\za_N=\cosh(Nu)\quad\rightarrow\quad \widetilde u=Nu,
\ee
one then easily arrives at
\be
z_n=\frac1{[2\sinh(nNu/2)]^D}.
\ee

\section{Discrete n-fermion energies}
\la{dneng}

The discrete $n$-fermion energies in 2D, for $n=4,5,6,10$, are given by
\ba
E_n(w)&=&\frac{{\cal N }_n(w)}{{\cal D }_n(w)}
\la{engn}
\ea
where
\ba
{\cal N }_4(w)&=&2 (99+196 \cosh(w)+140 \cosh(2 w)+96 \cosh(3 w)+33 \cosh(4 w)+12 \cosh(5 w))\nn\\
{\cal D}_4(w)&=&14 \sinh(w)+18 \sinh(2 w)+17 \sinh(3 w)+7 \sinh(4 w)+3 \sinh(5 w)\nn\\
{\cal N }_5(w)&=&4711+8989 \cosh(w)+7731 \cosh(2 w)+5999 \cosh(3 w)+4113 \cosh(4 w)\nn\\
&&\quad\  +2479 \cosh(5 w)+1245 \cosh(6 w) +532 \cosh(7 w)+168 \cosh(8 w)+33 \cosh(9 w)\nn\\
{\cal D}_5(w)&=&
117 \sinh(w)+197 \sinh(2 w)+223 \sinh(3 w)+197 \sinh(4 w)+143 \sinh(5 w)\nn\\
&&\quad +83 \sinh(6 w)+40 \sinh(7 w)+14 \sinh(8 w)+3 \sinh(9 w)\nn\\
{\cal N }_6(w)&=& 2 (13407+25721 \cosh(w)+23643 \cosh(2 w)+20003 \cosh(3 w)+16124 \cosh(4 w)
\nn\\ 
&&\quad+11811 \cosh(5 w) +8284 \cosh(6 w) +5100 \cosh(7 w)+2994 \cosh(8 w)+1495 \cosh(9 w)
\nn\\
&&\quad +697 \cosh(10 w)+231 \cosh(11 w)+83 \cosh(12 w)+7 \cosh(13 w))
\nn\\
{\cal D}_6(w)&=&319 \sinh(w)+586 \sinh(2 w)+736 \sinh(3 w)+783 \sinh(4 w)+707 \sinh(5 w)
\nn\\
&&\quad +588 \sinh(6 w) +415 \sinh(7 w)+275 \sinh(8 w)+152 \sinh(9 w)\nn\\
&&\quad +78 \sinh(10 w)+28 \sinh(11 w)+11 \sinh(12 w)+\sinh(13 w)\nn\\
{\cal N }_{10}(w)&=& 2 \bigl(2172390222+4314823362\cosh(w)+4226750929\cosh(2w)+4083387729\cosh(3w)
\nn\\ 
&&\quad+3890856869\cosh(4w)+3655779444\cosh(5w)+3387254600\cosh(6w)
\nn\\
&&\quad +3094022640\cosh(7w)+2786204840\cosh(8w)+2472642591\cosh(9w)
\nn\\
&&\quad +2162513611\cosh(10w)+1863006410\cosh(11w)+1580890927\cosh(12w)
\nn\\
&&\quad +1320648075\cosh(13w)+1085993727\cosh(14w)+878480571\cosh(15w)
\nn\\
&&\quad +698947301\cosh(16w)+546512238\cosh(17w)+419881330\cosh(18w)
\nn\\
&&\quad +316639653\cosh(19w)+234330752\cosh(20w)+169951624\cosh(21w)
\nn\\
&&\quad +120771356\cosh(22w)+83937273\cosh(23w)+57045201\cosh(24w)
\nn\\
&&\quad +37816435\cosh(25w)+24450854\cosh(26w)+15364872\cosh(27w)
\nn\\
&&\quad +9385562\cosh(28w)+5543351\cosh(29w)+3168210\cosh(30w)
\nn\\
&&\quad +1737590\cosh(31w)+916301\cosh(32w)+457947\cosh(33w)
\nn\\
&&\quad +218192\cosh(34w)+96260\cosh(35w)+39986\cosh(36w)+14646\cosh(37w)
\nn\\
&&\quad +4935\cosh(38w)+1274\cosh(39w)+295\cosh(40w)+15\cosh(41w)\bigr)\nn
\ea
\ba
{\cal D}_{10}(w)
&=&6269567\sinh(w)+12284154\sinh(2w)+17803408\sinh(3w)+22622674\sinh(4w)
\nn\\
&&\quad +26575827\sinh(5w)+29556940\sinh(6w)+31508496\sinh(7w)
\nn\\
&&\quad +32440172\sinh(8w)+32402745\sinh(9w)+31504002\sinh(10w)
\nn\\
&&\quad +29872385\sinh(11w)+27671748\sinh(12w)+25061386\sinh(13w)
\nn\\
&&\quad +22212164\sinh(14w)+19268782\sinh(15w)+16369539\sinh(16w)
\nn\\
&&\quad +13614501\sinh(17w)+11088828\sinh(18w)+8838635\sinh(19w)
\nn\\
&&\quad +6895528\sinh(20w)+5259530\sinh(21w)+3922450\sinh(22w)
\nn\\
&&\quad +2855501\sinh(23w)+2029305\sinh(24w)+1404510\sinh(25w)
\nn\\
&&\quad +946820\sinh(26w)+619536\sinh(27w)+393644\sinh(28w)+241576\sinh(29w)
\nn\\
&&\quad +143348\sinh(30w)+81552\sinh(31w)+44588\sinh(32w)+23086\sinh(33w)
\nn\\
&&\quad +11394\sinh(34w)+5203\sinh(35w)+2238\sinh(36w)+848\sinh(37w)
\nn\\
&&\quad +296\sinh(38w)+79\sinh(39w)+19\sinh(40w)+\sinh(41w)
\ea
In the continuum limit of $u\rightarrow \ep$, $w\rightarrow \tau$,
these are then the exact $n$-fermion energy as a function of $\tau$.

\section{Fourth-order algorithms}
\la{falg}

Positive time-step fourth-order short-time propagators with arbitrary number of 
$T$ operators, or beads, have been derived in Ref.\onlinecite{chin06}. For completeness, we will briefly
summarize their operator form as done in Ref.\onlinecite{chin15}.  
Consider a short-time propagator with $(N-1)$ beads of the form 
\ba
{\cal T}_{(N-1)B}^{(4)}(\epsilon)
&=&\e^{v_1\epsilon V}
\e^{t_2\epsilon T}\e^{v_2\epsilon V}
\cdots \e^{t_{N}\epsilon T}\e^{v_{N}\epsilon V}
\la{tb}
\ea
with left-right symmetric coefficients $v_1=v_{N}$, $t_2=t_{N}$, etc..
This operator will be fourth-order if one chooses $\{t_i\}>0$ satisfying
$\sum_{i=1}^{N}t_i=1$, and $\{v_i\}$ given by
\be
v_1=v_N=\frac12 + \lambda_2(1-t_2),\quad v_i=-\lambda_2(t_i+t_{i+1}),
\la{vcof}
\ee
where 
$\lambda_2=-\phi^{-1}/2$, $\ \phi=1-\sum_{i=1}^{N}t_i^3$, and the gradient coefficient $u_0=(1/\phi-1)/24$ with
the gradient potential term $u_0\epsilon^3[V,[T,V]]$ distributed
left-right symmetrically among all the $v_i\ep V$ terms in (\ref{tb}).
In order not to complicate the evaluation of Hamiltonian energy,
the gradient potential term must {\it not} be distributed to the $v_1$ and $v_N$ 
potential terms. 

To illustrate the above equations, consider the case of a 3-bead propagator with $N=4$.
One then must have $t_2+t_3+t_4=2t_2+t_3=1$ and therefore $t_3=1-2t_2$, with $t_2$ a free parameter
in the range $0\le t_2\le 1/2$.
One can then compute that $\phi=6t_2(1-t_2)^2$ and $u_0=(1/(6t_2(1-t_2)^2)-1)/24$,
reproducing the ${\cal T}_{BDA}$ propagator in Ref.\onlinecite{chin23a} with minor changes in notation.
The Hamiltonian energy can then be optimized by varying $t_2$ to give
the Best 3-Bead algorithm BB3. Its effectiveness in solving for the non-interacting fermion
energies is shown in Fig.\ref{n6opt}.

Similarly, one can derive the Best 4-Bead algorithm BB4 with the additional freedom of distributing the gradient term between the central $v_3V$ and the two adjacent $v_2V$ potential terms. For a BB5 algorithm, one has free parameters $t_2$, $t_3$ and the freedom to distribute the gradient term to the two inner or the two further out potential terms.

If the universal energy $E_n(w)$ is known, then any algorithm of the form (\ref{tb}), with given parameter values, can be contracted {\it numerically} down to $\ka_1$ and $\mu_1$ to determine $\ga(\ep)$ and therefore the Hamiltonian energy via (\ref{eh}), (\ref{hhw}) or (\ref{hww}).

\section{Average sign at large {\mbox{\boldmath $\tau$}}}
\la{ass}

To compute the average sign $s=\langle sgn \rangle$ for the two-fermion propagator in the three-bead case, one recalls
(\ref{tsign}). In the large $\tau$ limit, one can ignore all radial integration. For a closed loop one has
\be
s_3=\frac{\int_{-\pi}^{\pi}d\phi_1 \int_{-\pi}^{\pi}d\phi_2\cos(\phi_1)\cos(\phi_2)\cos(\phi_1+\phi_2)}
{\int_{-\pi}^{\pi}d\phi_1 \int_{-\pi}^{\pi}d\phi_2|\cos(\phi_1)\cos(\phi_2)\cos(\phi_1+\phi_2)|}
=\frac{I}{J}
\ee
which by symmetry
\be
I=\int_{0}^{\pi}d\phi_1 \int_{0}^{\pi}d\phi_2\cos(\phi_1)\cos(\phi_2)\cos(\phi_1+\phi_2)=\frac{\pi^2}{4}.
\ee
The integrand of $I$ has two negative triangular regions bounded by $\phi_1=\pi/2$, $\phi_2=\pi/2$, $\phi_1+\phi_2=\pi/2$,
and $\phi_1=\pi/2$, $\phi_2=\pi/2$, $\phi_1+\phi_2=3\pi/2$. Both integrated to $-(12-\pi^2)/32$. Hence 
\be
J=\frac{\pi^2}{4}+\frac{12-\pi^2}{8}=\frac{12+\pi^2}{8},
\ee
and
\be
s_3=\frac{2\pi^2}{12+\pi^2}=0.902586 .
\ee
For the general $N$-bead case,
\be
s_N=\frac{\int_{-\pi}^{\pi}d\phi_1 d\phi_2\cdots\phi_{N-1}\cos(\phi_1)\cos(\phi_2)\cdots\cos(\phi_{N-1})\cos(\sum_{k=1}^{N-1}\phi_k)}
{\int_{-\pi}^{\pi}d\phi_1 d\phi_2\cdots\phi_{N-1}|\cos(\phi_1)\cos(\phi_2)\cdots\cos(\phi_{N-1})\cos(\sum_{k=1}^{N-1}\phi_k)|},
\ee
the integral can be done numerically, either by quadrature (for small $N$) or by Monte Carlo (for larger $N$).
One then finds that
$s_4=0.7426$,
$s_5=0.5927$,
$s_6=0.4680$,
 and $s_8=0.2890$.

\section{No sign problem for $n=D+1$ fermions}
\la{signp}

By factoring out all positive factors in the 3-fermion free propagator from (\ref{gdet}), with $\s_i\equiv\br'_i$,
\ba
G_0(\s_i,\br_i;\ep)
&\propto&	\e^{-(\s^2_1+\s^2_2+\s^2_3+\br^2_1+\br^2_2+\br^2_3)/(2\ep)}
\det\left(\begin{array}{ccc}
	\e^{\s_1\cdot\br_1/\ep} & \e^{\s_1\cdot\br_2/\ep}& \e^{\s_1\cdot\br_3/\ep}\\
	\e^{\s_2\cdot\br_1/\ep} & \e^{\s_2\cdot\br_2/\ep}& \e^{\s_2\cdot\br_3/\ep}\\
	\e^{\s_3\cdot\br_1/\ep} & \e^{\s_3\cdot\br_2/\ep}& \e^{\s_3\cdot\br_3/\ep}
\end{array}\right), 
\la{threed}
\ea
its sign is given by the determinant above.
Multiply its first row by $\e^{(\s_2-\s_1)\cdot\br_1/\ep}\equiv\e^{\s_{21}\cdot\br_1/\ep}$ and $\e^{\s_{31}\cdot\br_1/\ep}$ 
and subtract them from the second and third row respectively gives
\ba
&&\det\left(\begin{array}{ccc}
	\e^{\s_1\cdot\br_1/\ep} & \e^{\s_1\cdot\br_2/\ep}& \e^{\s_1\cdot\br_3/\ep}\\
	0                  & \e^{\s_2\cdot\br_2/\ep}-\e^{\s_1\cdot\br_2/\ep+\s_{21}\cdot\br_1/\ep}&   
	\e^{\s_2\cdot\br_3/\ep}-\e^{\s_1\cdot\br_3/\ep+\s_{21}\cdot\br_1/\ep}\\
	0                  & \e^{\s_3\cdot\br_2/\ep}-\e^{\s_1\cdot\br_2/\ep+\s_{31}\cdot\br_1/\ep}& 
	\e^{\s_3\cdot\br_3/\ep}-\e^{\s_1\cdot\br_3/\ep+\s_{31}\cdot\br_1/\ep}
\end{array}\right),\nn\\
&=&\e^{\s_1\cdot\br_1/\ep}
\det\left(\begin{array}{cc}
	\e^{\s_2\cdot\br_2/\ep}(1-\e^{-\s_{21}\cdot\br_{21}/\ep})& 
	\e^{\s_2\cdot\br_3/\ep}(1-\e^{-\s_{21}\cdot\br_{31}/\ep})\\
	\e^{\s_3\cdot\br_2/\ep}(1-\e^{-\s_{31}\cdot\br_{21}/\ep})& 
	\e^{\s_3\cdot\br_3/\ep}(1-\e^{-\s_{31}\cdot\br_{31}/\ep}).
\end{array}\right).
\la{threep}
\ea
In the limit of $\ep\rightarrow \infty$, to leading order in $1/\ep$, the sign of the above determinant
is given by the expansion of each matrix element's bracket,
\ba
\det\left(\begin{array}{cc}
	(\s_{21}\cdot\br_{21})& 
	(\s_{21}\cdot\br_{31})\\
	(\s_{31}\cdot\br_{21})& 
	(\s_{31}\cdot\br_{31})
\end{array}\right)\equiv {\det}_3
&=& (\s_{21}\cdot\br_{21})(\s_{31}\cdot\br_{31}) -(\s_{21}\cdot\br_{31})(\s_{31}\cdot\br_{21}).
\la{tdot}
\ea
We will now evaluate the above directly in terms of the $D$ components of all the vectors
\ba
{\det}_3&=&\sum_{k_1=1}^D\sum_{k_2=1}^D\left[(\s_{21})_{k_1}(\br_{21})_{k_1} (\s_{31})_{k_2}(\br_{31})_{k_2}-(\s_{21})_{k_1}(\br_{31})_{k_1}(\s_{31})_{k_2}(\br_{21})_{k_2}\right]\nn\\
&=&\sum_{k_1=1}^D\sum_{k_2=1}^D(\s_{21})_{k_1}(\s_{31})_{k_2}
\det\left(\begin{array}{cc}
	(\br_{21})_{k_1}& 
	(\br_{31})_{k_1}\\
	(\br_{21})_{k_2}& 
	(\br_{31})_{k_2}
\end{array}\right)
\ea
Since the diagonal $k_1=k_2$ sum vanishes, one only needs to sum over
$k_1<k_2$ and $k_2<k_1$. The latter case is just the first case with
$k_1\leftrightarrow k_2$. Under this exchange the new determinant is just the old determinant with a negative sign, thereby giving
\ba
{\det}_3
&=&\sum_{1\le k_1<k_2\le D}\det\left(\begin{array}{cc}
	(\s_{21})_{k_1}& 
	(\s_{21})_{k_2}\\
	(\s_{31})_{k_1}& 
	(\s_{31})_{k_2}
\end{array}\right)
\det\left(\begin{array}{cc}
	(\br_{21})_{k_1}& 
	(\br_{31})_{k_1}\\
	(\br_{21})_{k_2}& 
	(\br_{31})_{k_2}
\end{array}\right)
\la{pro2}
\ea
For $D=2$, there is only a single term $(k_1=1, k_2=2)$ in the sum,
\ba
{\det}_3&=&[(\s_{21})_1(\s_{31})_2-(\s_{21})_2(\s_{31})_1][(\br_{21})_1(\br_{31})_2-(\br_{21})_2(\br_{31})_1]\nn\\
        &=& (\s_{21}\times\s_{31})_3(\br_{21}\times\br_{31})_3.
\ea
Therefore there is no sign problem. In 2D, the cross-product is the {\it signed area} of two relative vectors.

However, for $D=3$, there are three terms $(k_1=2,k_2=3)$, $(k_1=1,k_2=3)$, $(k_1=1,k_2=2)$ in the sum,
yielding
\ba
{\det}_3&=&[(\s_{21})_2(\s_{31})_3-(\s_{21})_3(\s_{31})_2][(\br_{21})_2(\br_{31})_3-(\br_{21})_3(\br_{31})_2]\nn\\
&+&[(\s_{21})_1(\s_{31})_3-(\s_{21})_3(\s_{31})_1][(\br_{21})_1(\br_{31})_3-(\br_{21})_3(\br_{31})_1]\nn\\
&+&[(\s_{21})_1(\s_{31})_2-(\s_{21})_2(\s_{31})_1][(\br_{21})_1(\br_{31})_2-(\br_{21})_2(\br_{31})_1]\nn\\
&=& (\s_{21}\times\s_{31})\cdot(\br_{21}\times\br_{31}).
\ea
The sum of three similar terms is a dot product, giving rise to the sign problem for three fermions in 3D.

For four fermions, similar manipulations as in the three-fermion case show that the sign is given by
\ba
{\det}_4=\det\left(\begin{array}{ccc}
	(\s_{21}\cdot\br_{21})&
	(\s_{21}\cdot\br_{31})& 
	(\s_{21}\cdot\br_{41})\\
	(\s_{31}\cdot\br_{21})& 
	(\s_{31}\cdot\br_{31})& 
	(\s_{31}\cdot\br_{41})\\
	(\s_{41}\cdot\br_{21})& 
	(\s_{41}\cdot\br_{31})& 
	(\s_{41}\cdot\br_{41})
\end{array}\right).
\la{pro4}
\ea
By again expanding out the components, one has 
\be
{\det}_4
=\sum_{{k_1}=1}^D\sum_{{k_2}=1}^D\sum_{{k_3}=1}^D(\s_{21})_{k_1}(\s_{31})_{k_2}(\s_{41})_{k_3}
\det\left(\begin{array}{ccc}
	(\br_{21})_{k_1}&
	(\br_{31})_{k_1}& 
	(\br_{41})_{k_1}\\
	(\br_{21})_{k_2}& 
	(\br_{31})_{k_2}& 
	(\br_{41})_{k_2}\\
	(\br_{21})_{k_3}& 
	(\br_{31})_{k_3}& 
	(\br_{41})_{k_3}
\end{array}\right).
\ee
Due to the determinant, the sum is non-vanishing only over ${k_1}\ne {k_2} \ne {k_3}$.
The sum now decomposes into six sums over permutations
$(k_1<k_2<k_3)$, 
$(k_1<k_3<k_2)$,
$(k_2<k_3<k_1)$,
$(k_2<k_1<k_3)$,
$(k_3<k_1<k_2)$, and
$(k_3<k_2<k_1)$.
By exchanging (relabeling) 2 indices at a time, the last five
sums can all be permuted back to the sum over $(k_1<k_2<k_3)$ having the same determinant as defined
in $(k_1<k_2<k_3)$ but with a positive sign for even number of exchanges and a negative sign for odd number of exchanges.
These sign changes then convert the product of three $\s$ relative vector's components into a determinant:
\be
{\det}_4=
\sum_{1\le k_1<k_2<k_3\le D}
\det\left(\begin{array}{ccc}
	(\s_{21})_{k_1}&
	(\s_{21})_{k_2}& 
	(\s_{21})_{k_3}\\
	(\s_{31})_{k_1}& 
	(\s_{31})_{k_2}& 
	(\s_{31})_{k_3}\\
	(\s_{41})_{k_1}& 
	(\s_{41})_{k_2}& 
	(\s_{41})_{k_3}
\end{array}\right)
\det\left(\begin{array}{ccc}
	(\br_{21})_{k_1}&
	(\br_{31})_{k_1}& 
	(\br_{41})_{k_1}\\
	(\br_{21})_{k_2}& 
	(\br_{31})_{k_2}& 
	(\br_{41})_{k_2}\\
	(\br_{21})_{k_3}& 
	(\br_{31})_{k_3}& 
	(\br_{41})_{k_3}
\end{array}\right).
\la{prop}
\ee
For $D=3$, there is only one term $(k_1=1,k_2=2,k_3=3)$ in the sum:
\ba
{\det}_4&=&
\det\left(\begin{array}{ccc}
	(\s_{21})_1&
	(\s_{21})_2& 
	(\s_{21})_3\\
	(\s_{31})_1& 
	(\s_{31})_2& 
	(\s_{31})_3\\
	(\s_{41})_1& 
	(\s_{41})_2& 
	(\s_{41})_3
\end{array}\right)
\det\left(\begin{array}{ccc}
	(\br_{21})_{1}&
	(\br_{31})_{1}& 
	(\br_{41})_{1}\\
	(\br_{21})_{2}& 
	(\br_{31})_{2}& 
	(\br_{41})_{2}\\
	(\br_{21})_{3}& 
	(\br_{31})_{3}& 
	(\br_{41})_{3}
\end{array}\right)\la{prop3}\\
&=&[\s_{21}\cdot(\s_{31}\times\s_{41})][\br_{21}\cdot(\br_{31}\times\br_{41})]
\nn
\ea
and therefore no sign problem. The vector triple product is well known to give the {\it signed volume} in 3D.
For $D>4$, $\det_4$ would have a sum of similar terms, which is a dot-product giving rise to a cosine function
and the sign problem.

Generalizing (\ref{pro4}) to the determinant of dot-products for $n$ fermions
can be written compactly as $\det(AB)$ where $A$ is a $(n-1)$ by $D$ matrix and $B$ is a $D$ by $(n-1)$ matrix given by
\be
A_{ij}=(\s_{(i+1)1})_j,\qquad B_{jk}=(\br_{(k+1)1})_j.
\ee 
The resulting decomposition of $\det(AB)$ into a sum of product of two determinants 
is then the well-known Cauchy–Binet formula\cite{wik}:
\be
\det(AB)=\sum_{1\le k_1<k_2<\cdots k_{n-1}\le D}\det(A)\det(B).
\la{cb}
\ee
When both $A$ and $B$ are square matrices, with $D=(n-1)$, then there is only
one term in the sum given by $\{k_i=i\}$, yielding the usual result of $\det(AB)=\det(A)\det(B)$,
which is a single number with no cosine function.
Therefore, in the large $\tau$ limit, there is no sign problem in $D$ dimension for $n=D+1$ fermions,
which is the first closed-shell state.
The two determinants $\det(A)$ and $\det(B)$ are then just $D$-dimension {\it hyper-volumes} formed by 
the two sets of relative vectors.

\end{document}